\documentclass[aps,prc,12pt,nofootinbib,superscriptaddress,showkeys,english]{revtex4-1}
\usepackage[latin1]{inputenc}
\usepackage[english]{babel}
\usepackage{graphicx}
\usepackage{color}
\usepackage{amsmath}
\usepackage{amssymb}
\usepackage{multirow}
\usepackage{color}

\begin{document}
\title{Magnetized vector boson gas at any temperature}
\author{G. Quintero Angulo}
\email{gquintero@fisica.uh.cu}
\affiliation{Facultad de F{\'i}sica, Universidad de la Habana,\\ San L{\'a}zaro y L, Vedado, La Habana 10400, Cuba}
\author{L. C. Su\'arez Gonz\'alez}
\email{lismary@icimaf.cu}
\affiliation{Instituto de Cibern\'{e}tica, Matem\'{a}tica y F\'{\i}sica (ICIMAF), \\
	Calle E esq a 15 Vedado 10400 La Habana Cuba}
\author{A. P\'erez Mart\'{\i}nez}
\email{aurora@icimaf.cu}
\affiliation{Departamento de F\'isica Fundamental, Universidad de Salamanca, \\
	Plaza Madrid s/n 37008, Salamanca, Espa\~na}
\affiliation{Instituto de Cibern\'{e}tica, Matem\'{a}tica y F\'{\i}sica (ICIMAF), \\
 Calle E esq a 15 Vedado 10400 La Habana Cuba}
\author{H. P\'erez Rojas}
\email{hugo@icimaf.cu}
\affiliation {Instituto de Cibern\'{e}tica, Matem\'{a}tica y F\'{\i}sica (ICIMAF), \\
 Calle E esq a 15 Vedado 10400 La Habana Cuba}

\begin{abstract}
We study the thermodynamic properties of a relativistic magnetized neutral vector boson gas at any temperature. By comparing the results with the low temperature and the non relativistic descriptions of this gas, we found that the fully relativistic case can be separated in two regimes according to temperature. For low temperatures, magnetic field effects dominate and the system shows a spontaneous magnetization, its pressure splits in two components and, eventually, a transversal magnetic collapse might occur. In the high temperature region, the gas behavior is led by pair production. The presence of antiparticles preserves the isotropy in the pressure, and increases the magnetization and the total pressure of the system by several orders. Astrophysical implications of those behaviors are discussed.

\end{abstract}


\maketitle

\section{Introduction}
\label{sec1}

The obtaining in the laboratory of the BEC for composite particles constituted one of the milestone of physics at the end of the last century, and it came along with the demonstration that Bose-Einstein condensation and superfluidity can also be considered as the limiting states of another more general phenomenon: fermion pairing \cite{Randeria,Leggett,Parish,116,118,123}.
Since the experimental achievement of Bose-Einstein condensation (BEC) \cite{Anderson1995}, bosonic gases have attracted lot of theoretical attention, not only for the condensate \emph{per se}, but also for other interesting  phenomena linked with it like the BEC collapse called 'Bose-Nova' \cite{donley2001dynamics}, the diffuse BEC of magnetized  charged gases \cite{ROJAS1996148,
ROJAS1997,Khalilov:1999,Khalilov1997,PEREZROJAS2000,Bargueno,Quintero2017IJMP} and the Bose-Einstein ferromagnetism \cite{yamada1982thermal,Simkin_1999}.

In Earth, BEC is still an exotic state restricted to lab, however astrophysical and cosmological environments provide appropriate conditions for its natural existence. Hypothetically, several types of mesons condensates --pions and kaons-- exist inside neutron stars \cite{Camezind,Lattimerprognosis,weber2017pulsars}. Moreover, in the last few years new observational evidence  has came to reinforce the supposition of a NS superfluid interior \cite{Page:2011yz,PageSC,Shternin:2010qi}. This superfluid is composed by paired protons and neutrons that, given the star´s typical inner conditions, are expected to be in an intermediate situation between the BCS and the BEC limits \cite{Gruber}. Usually, the paired nucleons are described in the BCS limit, although some descriptions in the BEC limit have also been developed giving birth to BEC stars models \cite{Chavanis:2011,PhysRevD.90.127501,Gruber,Quintero2018BECS}. Boson stars are less popular than fermion stars models to describe compact objects, nevertheless, self-gravitating boson systems have been studied since last century not only in connection to compact objects, but also as sources of dark mater and black holes~\cite{PhysRev.187.1767,PhysRevD.38.2376,Takasugi1984,Chavanis:2011}.

Besides dark matter stars, Bose-Einstein condensates are popular in cosmology as an alternative to the standard CDM model in small scales ($\sim 10$~kpc or less) \cite{hui2017ultralight}. Such models are based on the supposition that dark matter is composed of very light hypothetical bosons ($m \sim 10^{-21}-10^{-22}$~eV), known as axions \cite{hui2017ultralight,fukuyama2006relativistic,maleki2019deformed,das2018bose,calzetta2005early}.  Large-scale predictions of axion models are the same as in CDM, but small-scale predictions seem to be more in accordance with observations \cite{hui2017ultralight}. The BEC of axions has not only proven to be a viable dark matter candidate \cite{hui2017ultralight,maleki2019deformed}, but it also provides plausible explanations for dark energy and its relation with dark matter \cite{fukuyama2006relativistic,maleki2019deformed}.

A common approximation of all these theoretical studies is the assumption that the Bose gas is at zero or low temperature.  This is also the usual approximation in the case of fermion gases in astrophysical environments. In these scenarios, fermion densities are so high that thermal fluctuations become negligible even at the billions of kelvins reached inside neutron stars. However, this limit does not work as well for bosons, because due to BEC they are very sensitive to environmental changes (variations in particle density, temperature and magentic field) as we have already reported in some preliminary studies on magnetized spin-one particles at finite temperature~\cite{Lismary2018,de2019bose}.

Spin-one boson gases are of great interest due to their unique magnetic properties in connection with BEC \cite{yamada1982thermal} and with astrophysical magnetic field generation \cite{Quintero2018BECS}. All this background, makes us focus this article in the thermodynamic properties of a magnetized neutral vector boson gas (NVBG) at any temperature. In Section II and III we review the equations of motion of neutral vector bosons and the thermodynamic potential of the corresponding gas. Section IV is devoted to condensation, while Section V encloses the magnetic properties. In Section VI the equations of state (EoS) are discussed. Concluding remarks are listed in Section VII, while mathematical details are given in the appendix.

The numerical calculations and plots have been done for a composite spin-one boson formed by two paired neutrons, with mass $m=2m_N$ and magnetic moment $\kappa=2\mu_N$, being $m_N$ and $\mu_N$ the mass and the magnetic moment of the neutron. This kind of effective bosons might be created in the core of neutron stars (see \cite{Quintero2018BECS} and references therein). Apart from the astrophysical inspiration, the discussions of our results are valid for any massive neutral vector boson gas and could be applied to phenomena in condensate matter \cite{Simkin_1999} and heavy-ions colliders \cite{ayala1997density,begun2006particle,su2008thermodynamic}. Along the paper, the results for the relativistic vector boson gas at any temperature are compared with the ones coming for the low temperature \cite{Quintero2017PRC} and the non-relativistic \cite{Lismary2018} treatment of this gas. As we shall see, this provide a better understanding of the underlaying physics, as well as a quick way to detect the high temperature effects.

\section{Equation of motion of a neutral vector boson}
\label{sec2}

The Lagrangian of neutral vector bosons under the action of an external magnetic field is an extension of the original Proca Lagrangian for spin-one particles that includes particle-field interactions \cite{PhysRev.131.2326,PhysRevD.89.121701}
\begin{eqnarray}\label{Lagrangian}
\mathcal L = -\frac{1}{4}F_{\eta\nu}F^{\eta\nu}-\frac{1}{2} \psi^{\eta\nu}\psi_{\eta\nu}
+ m^2 \psi^{\eta}\psi_{\eta}
+i m \kappa(\psi^{\eta} \psi_{\nu}-\psi^{\nu}\psi_{\eta}) F_{\eta\nu},
\end{eqnarray}
\noindent where the index $\eta$ and $\nu$ run from 1 to 4, $F^{\eta\nu}$ is the electromagnetic tensor and  $\psi^{\eta\nu}$, $\psi^{\eta}$ are independent field variables that follow the equations of motion \cite{PhysRev.131.2326}
\begin{eqnarray}\label{fieldeqns}
\partial_{\eta} \psi^{\eta\nu}-m^2 \psi^{\nu}+ 2i \kappa m \psi^{\eta} {F_{\eta}}^{\nu}=0,\\
\psi^{\eta\nu} = \partial^{\eta} \psi^{\nu} - \partial^{\nu} \psi^{\eta},
\end{eqnarray}
\noindent that in the momentum space read \cite{Quintero2017PRC}
\begin{equation}
\left((p_{\eta} p^{\eta}  +  m^2)\delta_{\eta}^{\nu} -p^{\nu} p_{\eta}  - 2  i \kappa m {F_{\eta}}^{\nu}\right)\rho^{\eta} = 0.
\end{equation}
Thus, the vector boson propagator is
\begin{equation}\label{propagator}
D_{\eta\nu}^{-1}=(p_{\eta} p^{\eta}  +  m^2)\delta_{\eta}^{\nu} -p^{\nu} p_{\eta}  - 2  i \kappa m {F_{\eta}}^{\nu}.
\end{equation}
Taking the magnetic field uniform, constant and in $p_3$ direction $\textbf{B}=B\textbf{e}_3$, the generalized Sakata-Taketani Hamiltonian for the six component wave equation of the vector boson is obtained from Eq.~(\ref{fieldeqns}) \cite{PhysRev.131.2326, PhysRevD.89.121701}
\begin{eqnarray}\label{hamiltonian}
H = \sigma_3 m + (\sigma_3 + i \sigma_2) \frac{\textbf{p}^2}{2 m} -
i \sigma_2 \frac{(\textbf{p}\cdot\textbf{S})^2}{m}
-(\sigma_3 - i \sigma_2) \kappa \textbf{S} \cdot \textbf{B},
\end{eqnarray}
\noindent with $\textbf{p}=(p_{\perp},p_3)$ and $p_{\perp}=p_1^2 + p_{2}^2$. The $\sigma_{i}$ are the $2\times2$ Pauli
matrices and the $S_{i}$ are the $3\times3$ spin-one matrices in a representation in which $S_3$ is diagonal and $\textbf{S} = \{S_1,S_2,S_3\}$\footnote{
	$\begin{array}{ccc} S_1=\frac{1}{\sqrt{2}}
	\left( \begin{array}{ccc}
	0 & 1& 0\\
	1 & 0 & 1\\
	0 & 1 & 0
	\end{array}\right),
	& S_2=\frac{i}{\sqrt{2}}
	\left( \begin{array}{ccc}
	0 & \text{-}1& 0\\
	1 & 0 & \text{-}1\\
	0 & 1 & 0
	\end{array} \right),
	& S_3=
	\left( \begin{array}{ccc}
	1 & 0& 0\\
	0 & 0 & 0\\
	0 & 0 & \text{-}1\end{array}\right)\end{array}$}.

The spectrum of the bosons described by Eq.~(\ref{hamiltonian}) is
\begin{equation}
\varepsilon(p_3,p_{\perp}, B,s)=\sqrt{m^2+p_3^2+p_{\perp}^2-2\kappa s B\sqrt{p_{\perp}^2+m^2}},\label{spectrum}
\end{equation}
\noindent where $s=0, \pm 1$ are the spin eigenvalues.

Let us note that the magnetic field $B$ enters in the energy spectrum coupled with the transverse momentum component $p_{\perp}$ (see the last term in the previous equation). This coupling reflects the breaking of the $SO(3)$ symmetry of the free system and the axial symmetry imposed by the magnetic field. A difference with magnetized charged quantum particles is here the absence of Landau quantization in the transversal momentum component, a direct consequence of the electric neutrality of the bosons we are studying \cite{Quintero2017PRC}.

The ground state energy of the neutral vector bosons ($s=1$ and $p_3=p_{\perp}=0$) is
\begin{equation}
\varepsilon(0, b)=\sqrt{m^2-2\kappa B m}=m\sqrt{1-b},\label{massrest}
\end{equation}
\noindent  with $b=\frac{B}{B_c}$ and $B_c=\frac{m}{2\kappa}$. For the values of $m$ and $\kappa$ we are considering $B_c=7.8 \times 10^{19}~G$.

From Eq.~(\ref{massrest}) follows that the rest energy of the magnetized vector bosons decreases with the magnetic field and is zero for $B=B_c$. At this point the system becomes unstable \cite{Quintero2017PRC}. This instability is similar to the so-called zero-mode problem of magnetized charged spin-1 field. Thus, one expects that boson-boson interactions \cite{PhysRevD.52.7174} provide a mechanism to remove the instability through the condensation of vortices as in \cite{Ambjorn:1989sz}. However, in the present paper, we will neither deal with this phenomenon nor going beyond $B_c$, since the maximum magnetic field expected inside NSs is around $\simeq 5 \times 10^{18}$~G \cite{Lattimerprognosis}.

\section{Thermodynamic potential of the magnetized spin-one gas}
\label{sec3}

To obtain the thermodynamical potential of the magnetized NVBG we will follow the procedure showed in \cite{Quintero2017PRC}. We start from the spectrum Eq.~(\ref{spectrum}) and the definition
\begin{equation}\label{omega}
\Omega(B,\mu,T)= \Omega_{st}(B,\mu,T)+\Omega_{vac}(B),
\end{equation}
\noindent where
\begin{equation}
\Omega_{vac}(B)=\sum_{s}\int\limits_{0}^{\infty}\frac{p_{\perp}dp_{\perp}dp_3}{(2\pi)^2}\varepsilon(p_3,p_{\perp} B,s),
\end{equation}
is the zero-point energy or vacuum term and is only B-dependent. $\Omega_{vac}(B)$ has an ultraviolet divergence which can be easily treated since the theory of neutral vector bosons interacting with a magnetic field through the magnetic moment is renormalizable. Note in Eq.~(\ref{Lagrangian}) that the parameter $ m \kappa $, that plays the role of the coupling constant for the interactions of the bosons with the magnetic field, is dimensionless \cite{PhysRevC.99.065803,PhysRevC.101.035808,Zee2010,manohar2018introduction}. After renormalization (see Appendix A), the vacuum contribution reads
\begin{align}\label{Grand-Potential-vac}
\Omega_{vac}(b) = -\frac{m^4}{288 \pi}\left( b^2(66-5 b^2)
-3(6-2b-b^2)(1-b)^2 \log(1-b)\right.
\\\nonumber
\left. -3(6+2b-b^2)(1+b)^2\log(1+b) \right).
\end{align}

$\Omega_{st}$ is the statistical contribution of particles/antiparticles. It depends on the magnetic field intensity $B$, the chemical potential $\mu$ and the absolute temperature $T=1/\beta$, and can be written as
\begin{eqnarray}
\Omega_{st}(B,\mu,T)=  \sum_{s} \int\limits_{0}^{\infty}\frac{p_{\perp}dp_{\perp}dp_3}{(2\pi)^2 \beta} \ln \left( f_{BE}^{+} f_{BE}^{-}\right),
\end{eqnarray}
where $f_{BE}^{\pm} = \left [1-e^{-(\varepsilon\mp \mu)\beta} \right ] $ stands for particles/antiparticles.

To compute $\Omega_{st}$ we rewrite it as

\begin{equation}
\Omega_{st}(B,\mu,T)=  \sum_{s} \Omega_{st}(s),
\end{equation}
\noindent being $\Omega_{st}(s)$ the contribution of each spin state. Using the Taylor expansion of the logarithm, $\Omega_{st}(s)$ is transformed into
\begin{eqnarray}\label{Grand-Potential-sst1}
\Omega_{st}(s)= - \frac{1}{4 \pi^2 \beta}  \sum_{n=1}^{\infty} \frac{e^{n \mu \beta}+e^{- n \mu \beta }}{n}
 \int\limits_{0}^{\infty} p_{\perp}  dp_{\perp} \int\limits_{-\infty}^{\infty} dp_3 e^{-n \beta \varepsilon(p_3,p_{\perp}, B,s)},
\end{eqnarray}
\noindent where $e^{n \mu \beta}$ stands for the particles and $e^{- n \mu \beta}$ for the antiparticles.

After integration over $p_3$, partial integration over $p_{\perp}$ and the change of variables $x^2 = (m^2 + p_{\perp}^2 + \alpha^2)^2 - \alpha^2$, Eq.~(\ref{Grand-Potential-sst1}) becomes
\begin{eqnarray}\label{Grand-Potential-sst2}
\hspace{-0.5cm}\Omega_{st}(s)= - \frac{y_0^2}{2 \pi^2 \beta^2}  \sum_{n=1}^{\infty} \frac{e^{n \mu \beta}+e^{- n \mu \beta }}{n^2} K_2 (n \beta y_0)
- \frac{\alpha}{2 \pi^2 \beta}  \sum_{n=1}^{\infty} \frac{e^{n \mu \beta}+e^{- n \mu \beta }}{n} \int\limits_{y_0}^{\infty} dx
\frac{x^2  K_1 (n \beta x)}{\sqrt{x^2+\alpha^2}},
\end{eqnarray}
\noindent with $K_l(x)$ the McDonald function of order $l$, $y_0= m \sqrt{1-s b}$ and $\alpha=s m b/2$. Now $\Omega_{st}(b,\mu,T)$ reads
\begin{eqnarray}\label{Grand-Potential-sst21}
\hspace{-0.5cm}\Omega_{st}(b,\mu,T)= -  \sum_{s} \sum_{n=1}^{\infty} \frac{e^{n \mu \beta}+e^{- n \mu \beta }}{ 2 \pi^2 n \beta } \left \{ \frac{y_0^2}{n\beta^2} K_2 (n \beta y_0)
- \alpha \int\limits_{y_0}^{\infty} dx
\frac{x^2}{\sqrt{x^2+\alpha^2}} K_1 (n \beta x) \right \}.
\end{eqnarray}

We obtain the thermodynamic potential of the magnetized neutral vector boson gas at any temperature by adding  Eqs.~(\ref{Grand-Potential-vac}) and (\ref{Grand-Potential-sst21}). The thermodynamic magnitudes derived from Eqs.~(\ref{omega}--\ref{Grand-Potential-vac}--\ref{Grand-Potential-sst21}) will be study and compare with those that come from two important cases: the relativistic low temperature limit (LT) \cite{Quintero2017PRC} and the non--relativistic limit (NR) \cite{Lismary2018}.

The low temperature limit is obtained by assuming $T<<m$ and neglecting the antiparticles contribution as well as those of the spin states with $s=0,-1$ in Eq.~(\ref{Grand-Potential-sst21}). This last is equivalent to request, in addition, that $T<<2\kappa B<m$, since only for those temperatures transitions of bosons from the $s=1$ state to any excited spin state will be forbidden, meaning that this LT limit is also a strong field approximation. Further detail may be seen in Appendix B, where the computation of the thermodynamic magnitudes in the LT limit is sketched.

In the non--relativistic limit $p_3, p_{\perp}, \kappa B << m$. These approximations are equivalent to neglect the vacuum and the antiparticles contributions, and lead to the NR spectrum $\varepsilon(p,s)=m+\vec{p}^{\:2}/2m -s\kappa B$. Details of the computation of the NR thermodynamic quantities are shown in Appendix C.

According to the assumptions of the LT and the NR limits, to consider the magnetized NVBG at any temperature is equivalent to keep in Eqs.~(\ref{omega}--\ref{Grand-Potential-sst21}--\ref{Grand-Potential-vac}) the contributions of the antiparticles, as well as those of the vacuum and all the spin states.

For a Bose gas, the particle desnsity is
\begin{equation}\label{densitydefinition}
	\rho = \rho_{gs}-\frac{\partial \Omega}{\partial \mu},
\end{equation}
where $\rho_{gs}$ stands for the density in the ground state $\varepsilon(0,b)=m\sqrt{1-b}$  (the condensed ones), while the term $-\frac{\partial \Omega}{\partial \mu}=-\frac{\partial \Omega_{st}}{\partial \mu}$ accounts for the density in the excited states. In Eq.~(\ref{densitydefinition}), $\rho_{gs}$ is such that $\rho_{gs}=0$ for $T\geq T_c$, while $\rho_{gs}>0$ for $T<T_c$, being $T_c$ the critical temperature of condensation. Deriving with respect to the chemical potential in Eq.~(\ref{Grand-Potential-sst21}), we obtain the following expression for $\rho$ at any temperature
\begin{eqnarray}\label{density}
	\rho= \rho_{gs} + \sum_{s} \sum_{n=1}^{\infty} \frac{e^{n \mu \beta}-e^{- n \mu \beta}}{2 \pi^2} \left \{ \frac{y_0^2 }{n\beta} K_2 (n \beta y_0)  + \alpha \int\limits_{y_0}^{\infty} dx
	\frac{x^2}{\sqrt{x^2+\alpha^2}} K_1 (n \beta x) \right \}.
\end{eqnarray}

In Eq.~(\ref{density}), the particle density in the excited states can be written as the difference between the particles ($\rho^+$) and the antiparticles ($\rho^-$) in the system: $ \rho^+-\rho^-= -\frac{\partial \Omega}{\partial \mu}$. The particle/antiparticle density is thus obtained by taking only the terms with $e^{n \mu \beta}$ or $e^{-n \mu \beta}$ respectively.

\begin{figure}[ht!]
	\centering
	\includegraphics[width=0.6\linewidth]{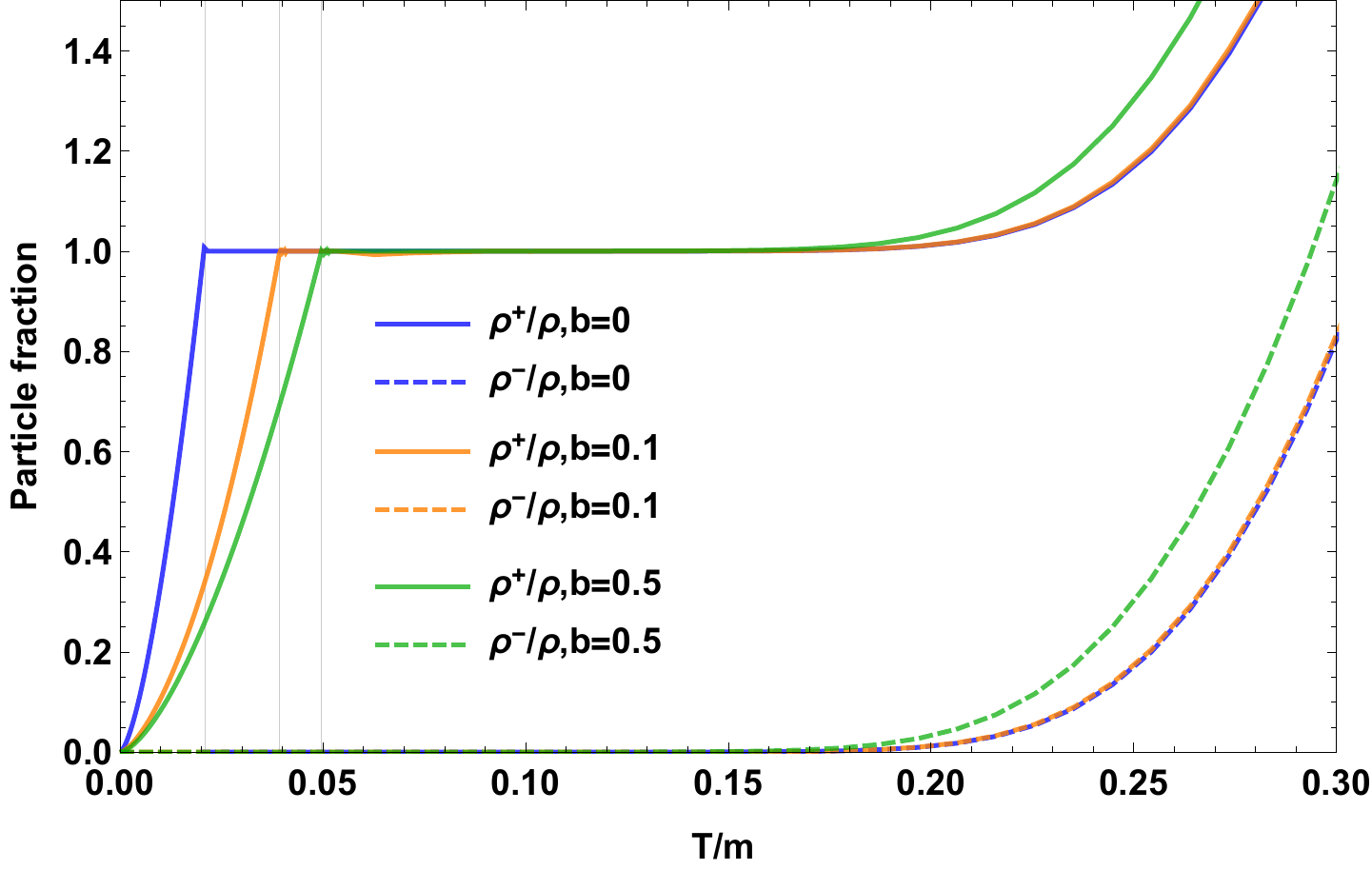}
	\caption{\label{f3.01} Particle (solid lines) and antiparticle (dashed lines) fraction as function of temperature for several values of the magnetic field. The horizontal lines indicate the BEC critical temperature $ T_c(b)$ (see Eq.~(\ref{criticalcurve}) in next section). }
\end{figure}

Fig.~\ref{f3.01} shows the fraction of non-condensed particles/antiparticles ($\rho^+/\rho $ and  $\rho^- / \rho$) as a function of the temperature and the magnetic field for $ \rho = 1.30 \times 10^{39}$cm$^{-3} $. The antiparticle density begins to be noticeable at $ T\gtrsim m/4 $ and increases with $B$. However, to appreciate this last effect ones requires magnetic fields close to $B_c$. Note that the curves for $ b = 0 $ and $ b = 0.1 $ are practically the same. For bosons with critical fields in the order of that of paired neutrons, the influence of $B$ in pair production is not relevant, but it could be important for particles with weaker critical fields.

\section{Bose--Einstein condensation}
\label{sec4}

Bose-Einstein condensation occurs when $\mu=m\sqrt{1-b}$ and $\rho_{gs}=0$ \cite{Pathria}. Setting this in Eq.~(\ref{density}) we get the following expression for the critical curve (i.e. for the implicit dependence of $\rho$, $T$ and $B$ in the transition points):
\begin{eqnarray}\label{criticalcurve}
\rho_c=  \sum_{s} \sum_{n=1}^{\infty} \frac{e^{n m \sqrt{1-b} \beta}-e^{- n m \sqrt{1-b} \beta}}{2 \pi^2} \left \{ \frac{y_0^2 }{n\beta} K_2 (n \beta y_0)  + \alpha \int\limits_{y_0}^{\infty} dx
\frac{x^2}{\sqrt{x^2+\alpha^2}} K_1 (n \beta x) \right \}.
\end{eqnarray}

\begin{figure}[ht!]
	\centering
	\includegraphics[width=0.49\linewidth]{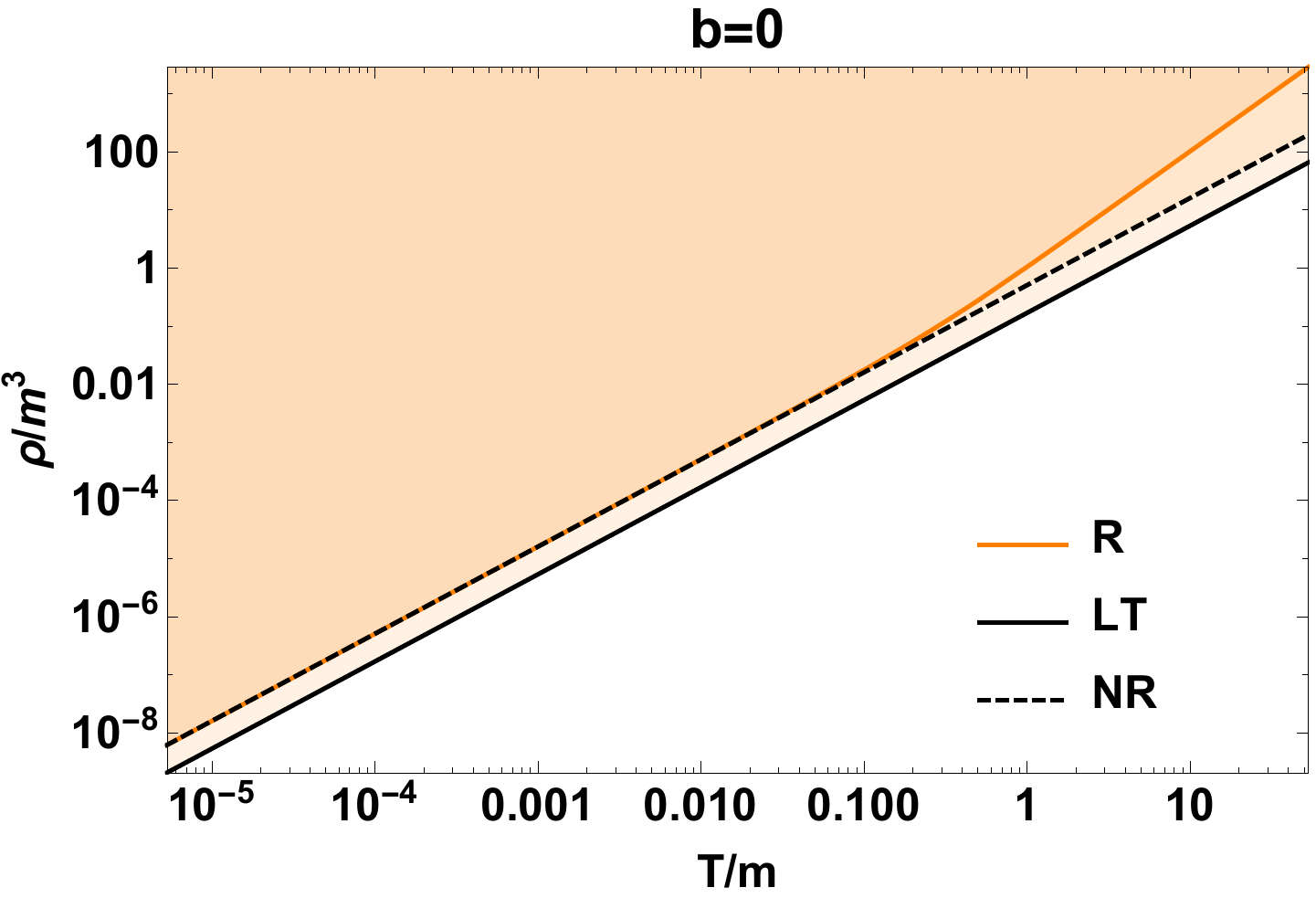}
	\includegraphics[width=0.49\linewidth]{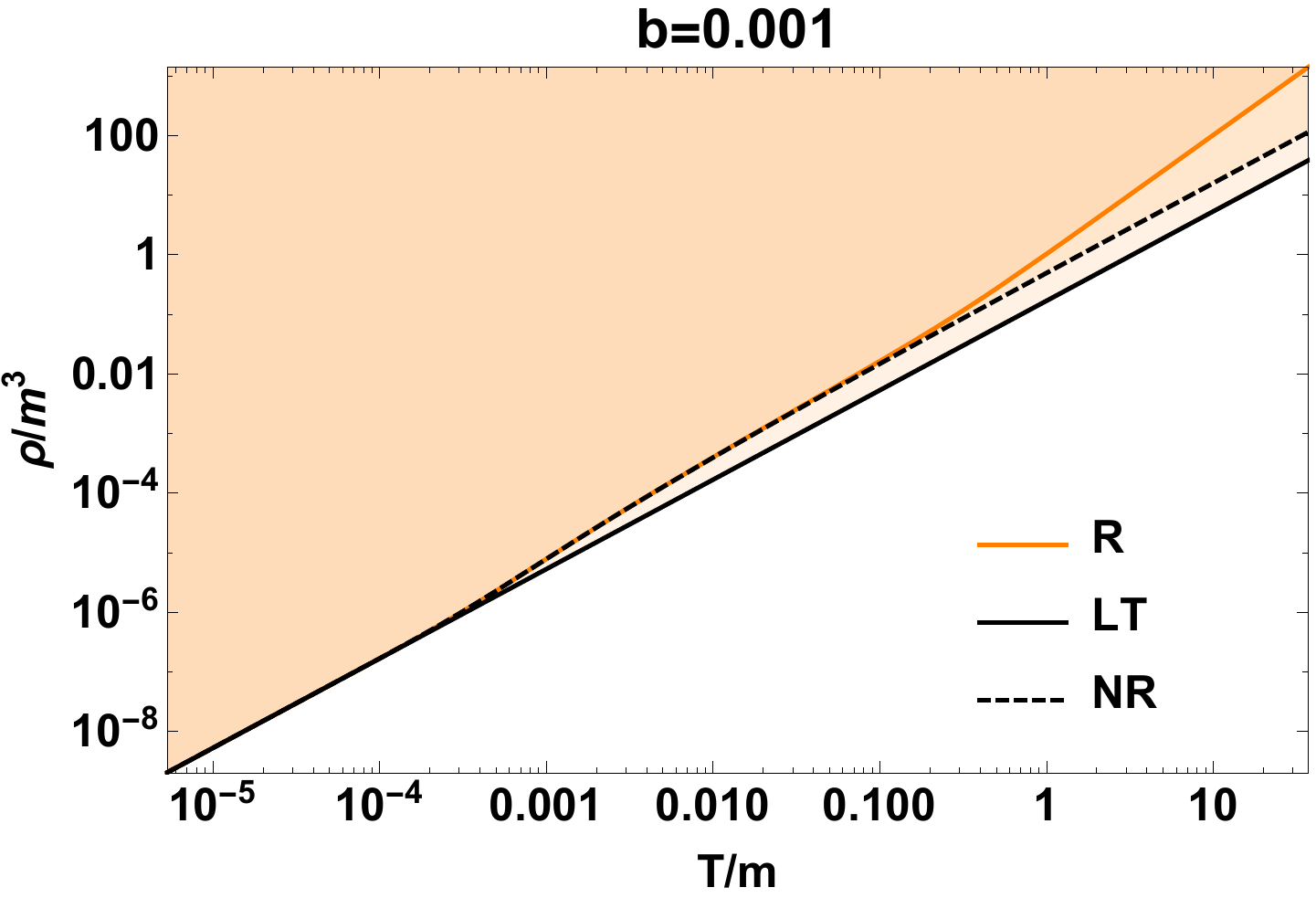}
	\caption{\label{F1} BEC phase diagram in the $\rho$ vs $T$ plane. The white region corresponds to the free gas while the colored one corresponds to the condensed state. The lines indicate the critical curves $ \rho_c(T, b)$ for the different descriptions of the NVBG.}
\end{figure}

\begin{figure}[ht!]
	\centering
	\includegraphics[width=0.6\linewidth]{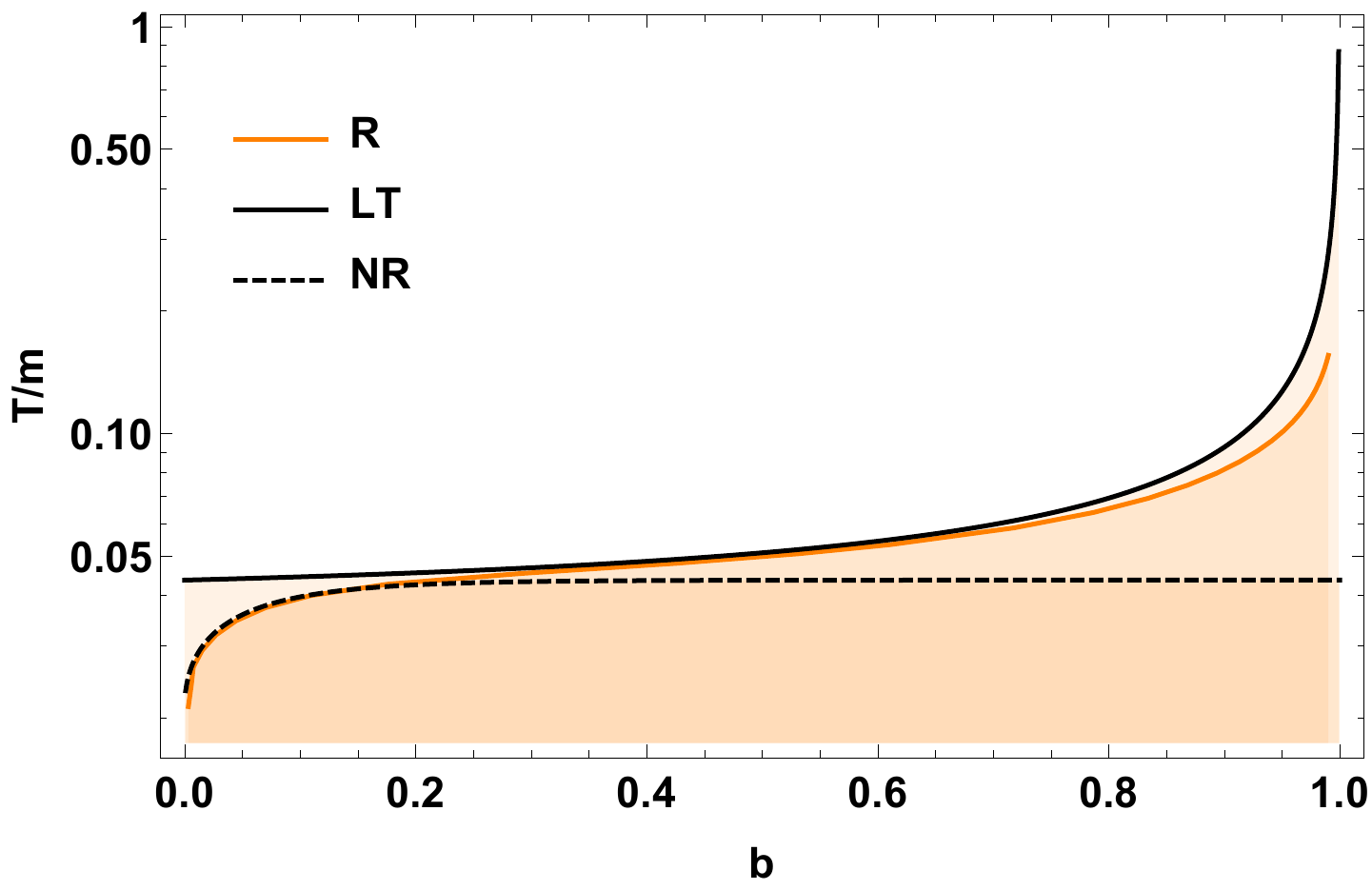}
	\caption{\label{F2} BEC phase diagram in the $T$ vs $B$ plane for $ \rho = 1.30 \times10^{39} cm^{-3} $. The white region corresponds to the free gas, and the colored one to the condensate.}
\end{figure}

Bose-Einstein condensation of magnetized Bose gases depends on three parameters: temperature, density and magnetic field; so the gas can reach the condensate in several ways. For instance, it condenses for fixed $\rho$ and $b$  when the temperature decreases; for fixed $T$ and $b$, when the density increases; and for fixed $\rho$ and $T$ if the magnetic field augments~\cite{Quintero2017PRC}.
We have illustrated these behaviors in Figs.~\ref{F1} and \ref{F2}, that correspond to the BEC phase diagrams in the  $\rho$ vs $T$ and the $T$ vs $b$ planes respectively.

Fig.~\ref{F1} shows the NVBG critical curves (Eq.~(\ref{criticalcurve})), denoted as R, along with the LT and the NR limits. Note that $\rho_c >> m^3$ at $T_c >> m$, which is the condition for a Bose gas to condense at relativistic temperatures~\cite{Roberts_2001}. In the low temperature region there is no difference in the behavior of the R and NR critical curves; they separate around $T \simeq m$ signaling the appearance of the antiparticles. On the other hand, from these plots is evident that the LT approximation is not valid in the non--magnetized case (remember it is also a strong field approximation). For  $b=0.001$, the LT critical curve coincides with the other two until $T \simeq 10^{-3} m$, indicating that this limit is not entirely correct above those temperatures.

In Fig.~\ref{F2} we draw the BEC phase diagram in the $T$ vs $b$ plane. As it is shown, the increasing of $b$ augments $T_c$ in the relativistic cases, and, as $b \rightarrow 1$($B\rightarrow B_c$) the critical temperature of the relativistic gases diverge, while in the NR limit it approaches the constant value $T^{NR}_{c}(\infty)=\frac{2\pi}{m} \left(\rho/ \zeta(3/2) \right)^{2/3}$, where $\zeta(x)$ is the Riemann zeta function \cite{Lismary2018}. The saturation of $T^{NR}_c(b)$ is caused by the absence of a critical magnetic field. For magnetic fields before saturation, we find that increasing $B$ increases $T^{NR}_c$ in an noticeable way, driving the system to condensation. But when the magnetic field reaches the saturated region, further changes on it barely affects $T^{NR}_c$.

The divergence of the critical temperature of the relativistic gases when  $b \rightarrow 1$ means that the gas is always condensed, regardless its density. This can also be seeing if we use Eq.~(\ref{Tclt}) to compute the critical density $\rho^{LT}_c(T,b)$ in the LT limit. The result is
\begin{equation}
\rho^{LT}_c(T,b)=\frac{\zeta(3/2)}{\sqrt{2}(2-b)} \left (\frac{m \sqrt{1-b}}{\pi \beta}\right)^{3/2}.
\end{equation}
From the above expression it is easily seen that $\rho^{LT}_c(T,c)=0$ for $b=1$. Since the extension of our calculations to $b>1$ is not straightforward \cite{Khalilov:1999}, no conclusions can be made about this region.

The enhancing effect of the magnetic field on BEC is related to the way it modifies the ground state of the NVBG. In general, the critical temperature of BEC depends on the inverse of the rest mass  of the bosons $\varepsilon_{gs}$, so that $T_c \rightarrow \infty$ as $\varepsilon_{gs} \rightarrow 0$, i.e., decreasing  $\varepsilon_{gs}$ favors BEC \cite{PhysRevLett.46.1497,ROJAS1997}. For the magnetized NVBG, $\varepsilon_{gs}=\varepsilon(0,b)=m\sqrt{1-b}$, and as $b$ increases, $\varepsilon(0,b)$ decreases, augmenting $T_c$ and driving the system to the condensate. A similar enhancing effect on BEC due to the magnetic field has been found and discussed in \cite{Bargueno} and \cite{Simkin_1999}.

It is also worth noting in Fig.~\ref{F2} that at $b=0$ the R and NR critical temperatures coincide, $T_c(0) = T^{NR}_c(0) = \frac{2\pi}{m} \left( \frac{\rho}{3 \zeta(3/2)}  \right)^{2/3}$, while $ T^{LT}_c(0) = \frac{2\pi}{m} \left( \frac{\rho}{\zeta(3/2)}  \right)^{2/3}$. This difference arises because in the LT limit, the spin states with $s=0,-1$ were neglected and all the particles are considered to have $s=1$. Therefore, once this approximation is done, the $b=0$ case can not be recovered. This is in agreement with  Fig.~\ref{F1} and highlights that the our LT limit is not suitable for weak magnetic fields. Finally, let us note that $ T^{LT}_c(0) = T^{NR}_c(\infty)$, since in the NR limit, $b \rightarrow \infty$ drives the system to a state in which all the particles are aligned with the magnetic field, i.e., they all are in the $s=1$ state.

\section{Magnetic properties}
\label{sec5}

In this section we focus on the dependence of the magnetization of the gas on the temperature and the magnetic field. The explicit analytical form of this dependence is derived from the definition
\begin{eqnarray}
\mathcal M &=&\frac{\kappa}{\sqrt{1-b}} \rho_{gs}-\frac{\partial \Omega_{st}}{\partial B}
-\frac{\partial \Omega_{vac}}{\partial B}\label{magdef}.
\end{eqnarray}
The first term in Eq.~(\ref{magdef}), ${\mathcal M}_{gs}=\frac{\kappa}{\sqrt{1-b}} \rho_{gs}$, stands for the magnetization of the condensed particles. It has to be added because all of the condensed bosons are aligned to the field, but when the condensate is present $\Omega_{st}$ only accounts for the particles in the excited states. The other two terms corresponds to the magnetization of the free particles ${\mathcal M}_{st}=-\frac{\partial \Omega_{st}}{\partial B}$ and the vacuum ${\mathcal M}_{vac}=-\frac{\partial \Omega_{vac}}{\partial B}$. They read
\begin{eqnarray}\label{magst}
{\mathcal M}_{st} = \sum_{s} \frac{\kappa s}{\pi^2 \beta } \sum_{n=1}^{\infty}  \frac{e^{n \mu \beta}+e^{- n \mu \beta }}{n} \left \{ \frac{m y_0 }{ (2-b s)} K_1 (n \beta y_0) +\int\limits_{y_0}^{\infty} dx
\frac{x^4}{ 2 (x^2+\alpha^2)^{3/2}} K_1 (n \beta x) \right \},  \quad \quad
\end{eqnarray}
and
\begin{eqnarray}\label{magvac}
{\mathcal M}_{vac}=-\frac{\kappa m^3}{72 \pi} \left \{ 7 b (b^2-6) -3(2b^3-9b+7)\log(1-b) -3(2b^3-9b-7)\log(1+b)\right \}. \quad \quad
\end{eqnarray}

\begin{figure}[ht!]
	\centering
	\includegraphics[width=0.49\linewidth]{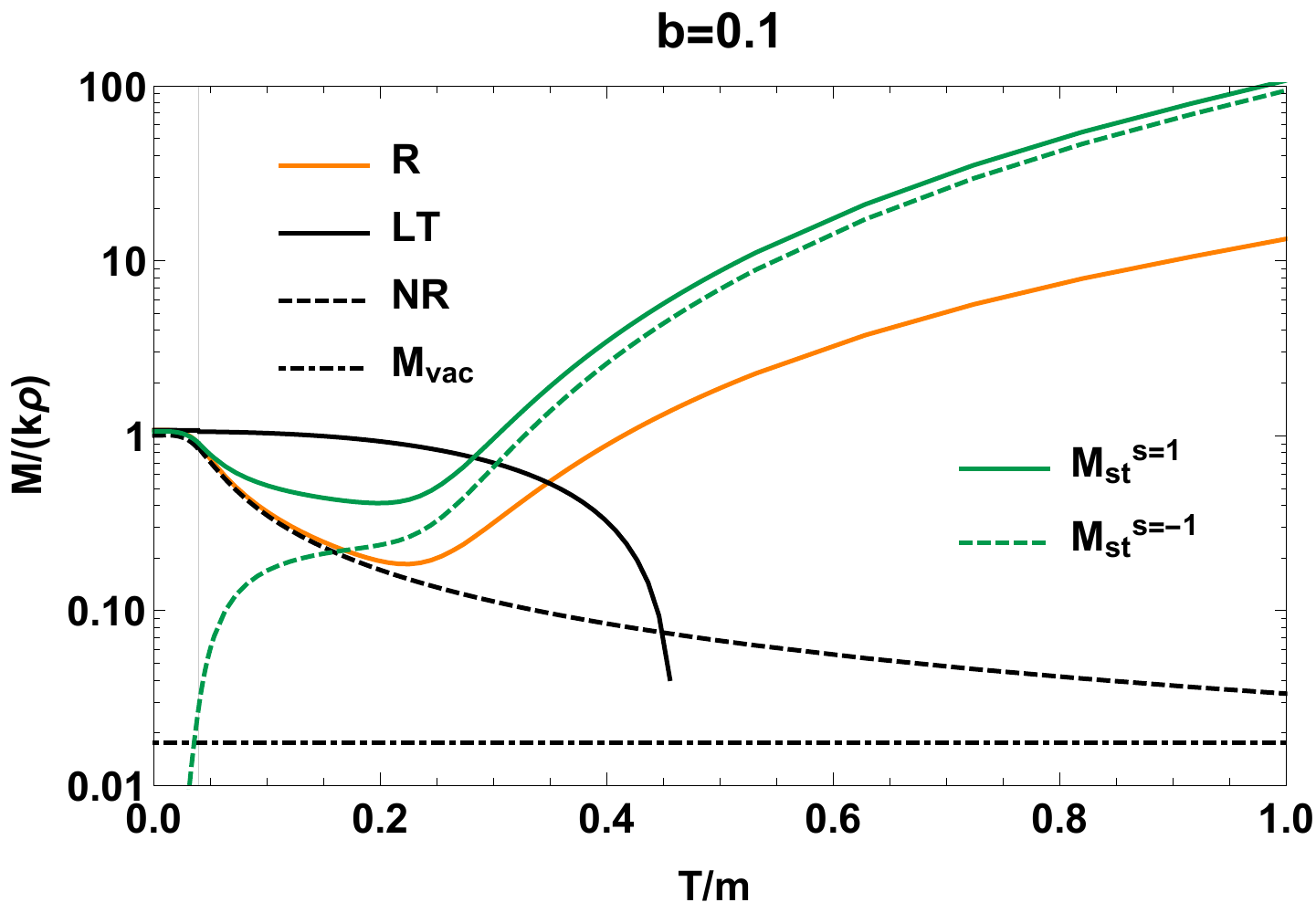}
	\includegraphics[width=0.49\linewidth]{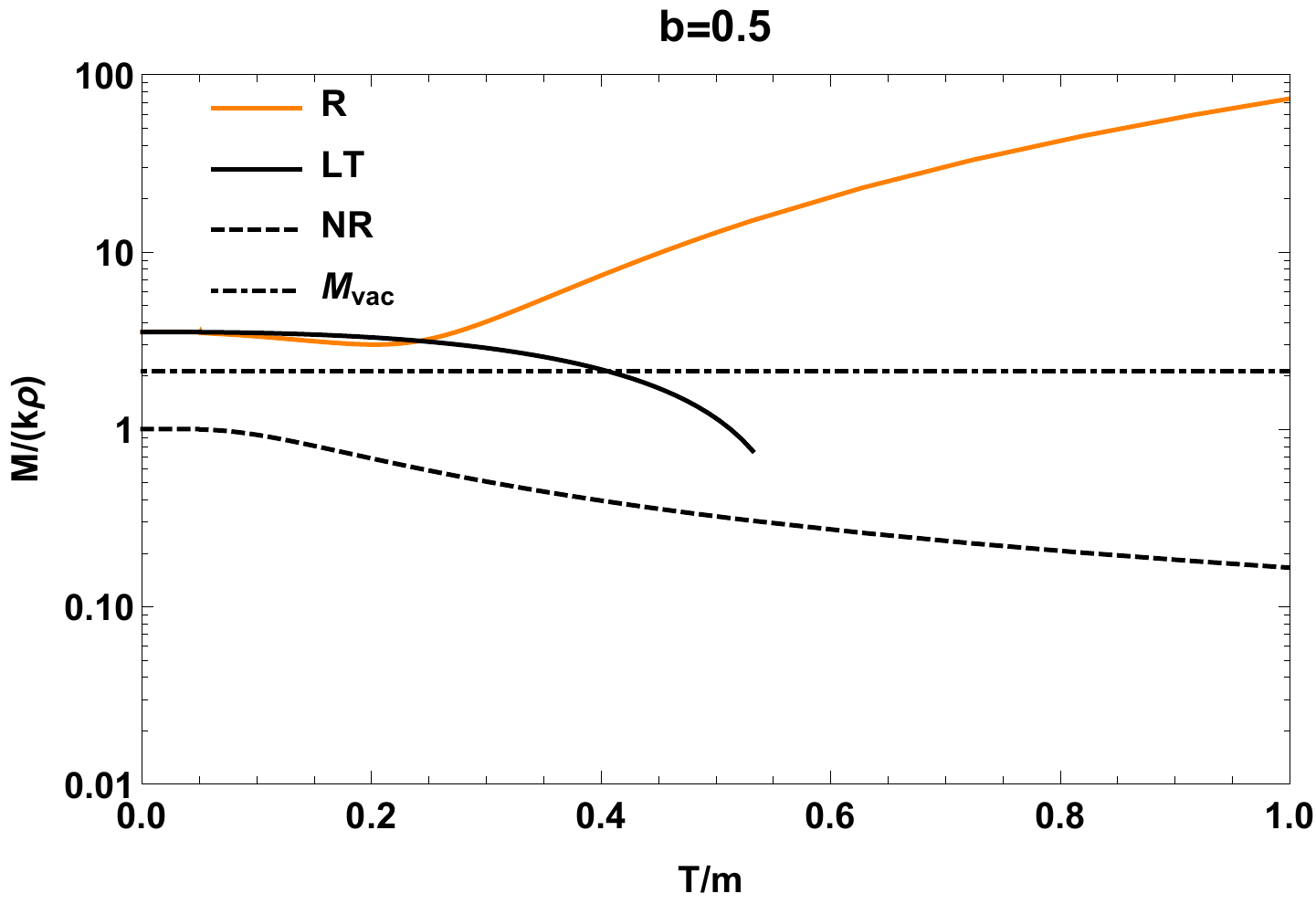}
	\caption{\label{mag} The magnetization as a function of temperature for $ \rho=1.30 \times 10^{39}$ cm $^{-3} $. The vertical line in left panel marks the temperature of BEC.}
\end{figure}

Fig.~\ref{mag} shows the total magnetization of the gas as a function of temperature for $ \rho=1.30\times10^{39}$cm$^{-3} $ and two values of $b$, $0.1$ and $0.5$. The LT and NR limits were drawn for comparison, as well as ${\mathcal M}_{vac}$. To facilitate the discussion, in the left panel we also plot the absolute value of the magnetization per spin state $M_{st}^{s=1}$ and $M_{st}^{s=-1}$, i.e., the expression under the sum over the spin states in Eq.~(\ref{magst}) evaluated for $s=\pm1$ ($M_{st}^{s=0}$ is zero).

For $b=0.1 $ the vacuum magnetization is negligible, and the R, NR and LT curves coincide for $ T\rightarrow0 $ and tend to $\kappa \rho $. As the temperature increases, the NR magnetization decreases and goes to zero for $T \rightarrow \infty$ \cite{Lismary2018}. The magnetization of the relativistic gas behaves like that of the non-relativistic limit up to $ T \sim 0.2\; m $. After this temperature, $\mathcal{M} (\mu, T, b)$ begins to grow and increases in several orders. This counterintuitive result stems from the quantum-relativistic character of the problem we are studying. Two factors jointly contribute to the increment of the magnetization with the temperature: 
\begin{enumerate}
	\item Particles and antiparticles with opposite spin couple differently to the magnetic field. 
	
	Note that the relation between the effective magnetic moment $d(s)=\frac{\partial \varepsilon(p_3,p_{\perp},B,s)}{\partial B}$ of each spin state is
	\begin{equation}\label{effectivemagneticmoment}
		d(1)=\frac{\kappa \sqrt{p_{\perp}^2+m^2}}{\sqrt{m^2+p_3^2+p_{\perp}^2-2\kappa B\sqrt{p_{\perp}^2+m^2}}} > \frac{\kappa \sqrt{p_{\perp}^2+m^2}}{\sqrt{m^2+p_3^2+p_{\perp}^2+2\kappa B\sqrt{p_{\perp}^2+m^2}}} = |d(-1)|.
	\end{equation}	
	Hence, particles (antiparticles) in opposite spin states do not contribute to the magnetization on equal footing. The contribution of particles (antiparticles) in the $s=1$ state is bigger.	
	
	\item The particle and antiparticle densities increase with temperature (Fig.~\ref{f3.01}).
	
\end{enumerate}
The former factor is derived from the relativistic energy expression Eq.~(\ref{spectrum}), whereas the latter is a consequence of pair production (an inherently quantum-relativistic effect). To better understand how they work together, let us focus on the $M_{st}^{s=1}$ and $M_{st}^{s=-1}$ curves in the left panel of Fig.~\ref{mag}. In the high temperature region, the contribution of the particles with $s=-1$ to the magnetization is significant and increases with $T$ due to pair production. But, as Eq.~(\ref{effectivemagneticmoment}) indicates, $M_{st}^{s=-1}$ is always smaller than $M_{st}^{s=1}$ resulting in a total magnetization, $M=M_{st}^{s=1}-M_{st}^{s=-1}$, that is always positive and also increases with the temperature\footnote{Above the condensation temperature, $M_{vac}$ is negligible and $M_{gs}=0$.}. This is consistent with the fact that in the $T \rightarrow \infty$ limit, $M_{st}$ (Eq.~(\ref{magst})) tends to infinity regardless the value of the magnetic field.

Both panels of Fig.~\ref{mag} show that the LT magnetization also decreases when $T$ increases, but its behavior is quite different from the other two cases, becoming negative around $T \sim0.5 \; m$ (the point where the curve ends). However, this negative magnetization does not imply the gas having a diamagnetic behavior; it is again a consequence of neglecting the states with $s = 0$ and $s = -1$ in the LT limit. The behavior of the magnetization in this limit reinforces the fact that it is only valid for $T \lesssim10^{-3} \; m$, something that can be also appreciated in the right panel of Fig.~\ref{F1}.

For $b=0.5$, $\mathcal M_ {vac}$ is higher than the maximum of $\mathcal{M}^{NR}$ and comparable to $\mathcal M_ {st}$. As a consequence, the magnetization of the relativistic cases differ from $\mathcal{M}^{NR}$ at $T = 0$. Since the LT limit works better for strong magnetic field, for $b=0.5$ the LT and the R magnetization curves coincide in a larger interval of temperature.

Fig.~\ref{mag} highlights the importance of considering the effects of antiparticles and the vacuum, which are usually neglected. In particular, in the case of antiparticles, they begin to be relevant for $ T \sim 0.25 \; m $, which for bosons formed by two neutrons is equivalent to $ T \sim10^{12} $ K, a relatively high temperature for astrophysical environments. But if we consider a lighter particle, such as positronium, $T \sim 0.25 \; m $ equals $ T\sim10^{9} $ K, a temperature achievable in the early stages of neutron star life.

\subsection{Bose-Einstein ferromagnetism}

It is also interesting to analyze the limit $b \rightarrow 0 $ in Eq.~(\ref{magdef}). $\mathcal M_{vac} (b=0)= 0$, while setting $ b = 0 $ in Eq. (\ref{magst}) gives
\begin{eqnarray}\label{eq3.10}
\mathcal{M}_{st}=\sum_{s} \frac{m^2 \kappa s}{2 \pi^2 \beta} \sum_{n=1}^{\infty} \!\frac{z^n+z^{-n}}{n}\bigg\{K_1(nm/T)+\int_{m}^{\infty}xK_1(nx/T)\;dx\bigg\},
\end{eqnarray}
\noindent but if we sum by $ s =\pm 1 $ in the previous expression it also equals zero. However, below $T_c$, $\rho_{gs} (T) \neq 0$, and the magnetization is different from zero even if $ b=0 $
\begin{equation}\label{eq3.11}
\mathcal{M}^{\pm}(\mu,T,0)= \kappa \rho_{gs} (T).
\end{equation}

Eq.~(\ref{eq3.11}) demonstrates that a spin-one BEC that was under the action of an external magnetic field, will remain magnetized even if the external magnetic field is somehow ``disconnected'' \cite{Quintero2017PRC,Lismary2018}. This phenomenon, known as Bose-Einstein ferromagnetism \cite{yamada1982thermal}, is a consequence of BEC, since all the bosons in the ground state have $s = 1$ (see Eq.~\ref{massrest}). To check out the connection between the magnetic behavior of the gas and the Bose--Einstein condensation, we will look at the specific heat and the magnetic susceptibility, whose maximum signal the corresponding phase transitions.

To compute the specific heat $C_v = \partial E/\partial T$, we need the internal energy density
\begin{equation}\label{energdef}
E=\Omega-T \frac{\partial \Omega}{\partial T} - \mu \frac{\partial \Omega}{\partial \mu}.
\end{equation}

After derivation of the thermodynamical potential with respect to the temperature and some simplifications, we find the entropy of the gas $S = -\partial \Omega/\partial T $ to be
\begin{eqnarray}\label{entropy}\nonumber
S =- \frac{\mu}{T}(\rho^+-\rho^-)-\frac{2}{T}\Omega_{st}
+ \sum_{s}  \sum_{n=1}^{\infty}  \frac{e^{n \mu \beta}+e^{- n \mu \beta }}{n} \left \{ \frac{ y_0^3 }{4 \pi^2}[ K_1 (n \beta y_0)+K_3 (n \beta y_0)] \right. \nonumber \\
+ \frac{\alpha n}{2 \pi^2 T}\left. \int\limits_{y_0}^{\infty} dx
\frac{x^3}{ \sqrt{x^2+\alpha^2}} K_0 (n \beta x) \right\},
\end{eqnarray}
\noindent and combining Eqs.(\ref{density}), (\ref{entropy}) and (\ref{energdef}), the internal energy can be written as
\begin{eqnarray}\label{energy}
E =-\Omega_{st} +\Omega_{vac}
+\sum_{s}  \sum_{n=1}^{\infty}  \frac{e^{n \mu \beta}+e^{- n \mu \beta }}{n}  \left \{ \frac{ y_0^3 T }{4 \pi^2}[ K_1 (n \beta y_0)+K_3 (n \beta y_0)] \right. \nonumber \\\nonumber
+ \frac{\alpha n}{2 \pi^2}\left. \int\limits_{y_0}^{\infty} dx
\frac{x^3}{ \sqrt{x^2+\alpha^2}} K_0 (n \beta x) \right \},
\end{eqnarray}
\noindent while the specific heat is
\begin{eqnarray}\label{cvT}
C_v &=& S+\sum_{s}\sum_{n=1}^{\infty}\bigg\{\frac{y_0^3 (z^n+z^{-n})}{4\pi^2 n}[K_1(n\beta y_0)+K_3(n \beta y_0)]\\\nonumber
&+&\frac{y_0^4 (z^n+z^{-n})}{8 \pi^2T}[K_1(n \beta y_0)+2K_2(n\beta y_0)+K_4(n\beta y_0)]\\\nonumber
&-&\frac{\mu (z^n-z^{-n})}{T}\bigg[\frac{y_0^3 (K_1(n\beta y_0)+K_3(n\beta y_0))}{4\pi^2}
+\frac{\alpha n}{2\pi^2T}\int_{y_0}^{\infty}\frac{x^3 K_0(n \beta x)dx}{\sqrt{x^2+\alpha^2}}\bigg]\\
&+&\frac{\alpha n (z^n+z^{-n})}{2\pi^2T^2}
\int_{y_0}^{\infty}\frac{x^4 K_1(n\beta x)dx}{\sqrt{x^2+\alpha^2}}\bigg\}.\nonumber
\end{eqnarray}

The magnetic susceptibility $\chi = -\partial \mathcal M/\partial B$ turns out to be
\begin{eqnarray}\label{chi}
\chi &=&\left\{
\begin{array}{cc}
\chi^{T>T_c}+\chi^{vac}, &\text{Free gas} \vspace{10pt} \\
\chi^{T<T_c}+\chi^{vac}, & \text {BEC}
\end{array}\right.
\end{eqnarray}
\noindent where
\begin{equation}\nonumber
	\chi^{T>T_c} = \frac{\partial {\mathcal M}_{st}}{\partial B},
\end{equation}
\begin{eqnarray}\label{chiT}
\chi^{T>T_c} = \sum_{s}\sum_{n=1}^{\infty} \frac{\kappa^2 s^2 (z^n-z^{-n})}{\pi^2 n}\bigg\{   \frac{m^2 n}{(2-bs)}K_0(n\beta y_0)
+\frac{2(4-3bs) m T}{(2-bs)^3}K_1(n\beta y_0)\\-\frac{3\alpha s T}{2}\int_{y_0}^{\infty}\frac{x^4K_1(n\beta x)dx}{(x^2+\alpha^2)^{5/2}}\bigg\},\nonumber
\end{eqnarray}
\noindent and
\begin{equation}\nonumber
\chi^{T<T_c}= \chi^{T>T_c}+\frac{\partial {\mathcal M}_{gs}}{\partial B},
\end{equation}
\noindent with
\begin{equation}\label{chiTc}\nonumber
\frac{\partial {\mathcal M}_{gs}}{\partial B}=\frac{\kappa^2 \rho_{gs}}{m\sqrt{(1-b)^3}}
-\sum_{s}\sum_{n=1}^{\infty} \frac{k^2 s (z^n-z^{-n})}{\pi^2 \sqrt{1-b}} \bigg\{\frac{m y_0}{(2-bs)}K_1(n\beta y_0)
+\frac{1}{2}\int_{y_0}^{\infty}\frac{x^4K_1(n\beta x)dx}{(x^2+\alpha^2)^{3/2}} \bigg\},
\end{equation}
\noindent and
\begin{eqnarray}\label{chivac}\nonumber
\chi_{vac}= \frac{\partial {\mathcal M}_{vac}}{\partial B}
= \frac{\kappa^2 m^2}{4 \pi} \left\{ -b^2 +(2b^2-3)\log(1-b^2)  \right\}.
\end{eqnarray}

\begin{figure}[ht!]
	\centering
	\includegraphics[width=0.49\linewidth]{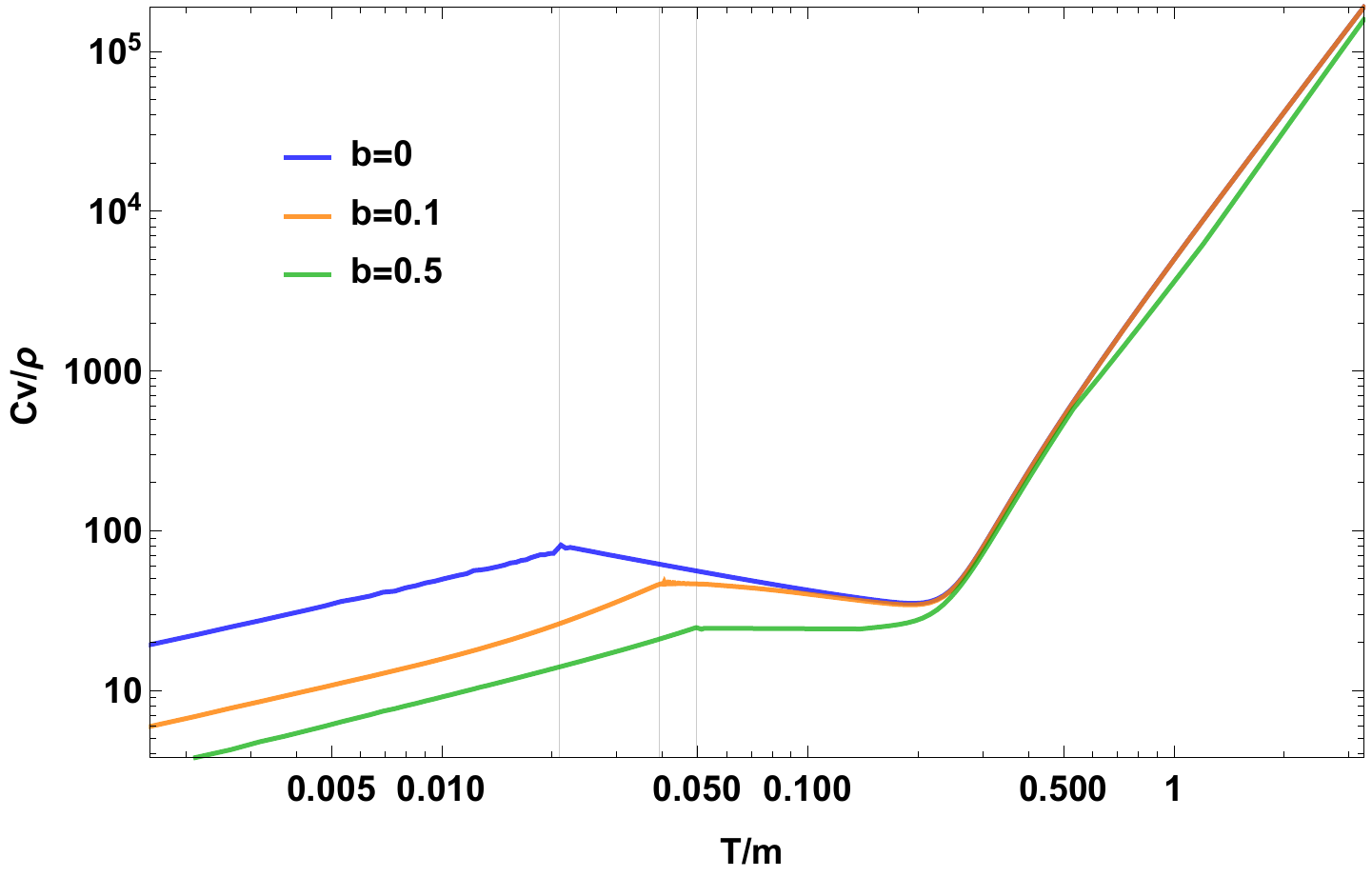}
	\includegraphics[width=0.49\linewidth]{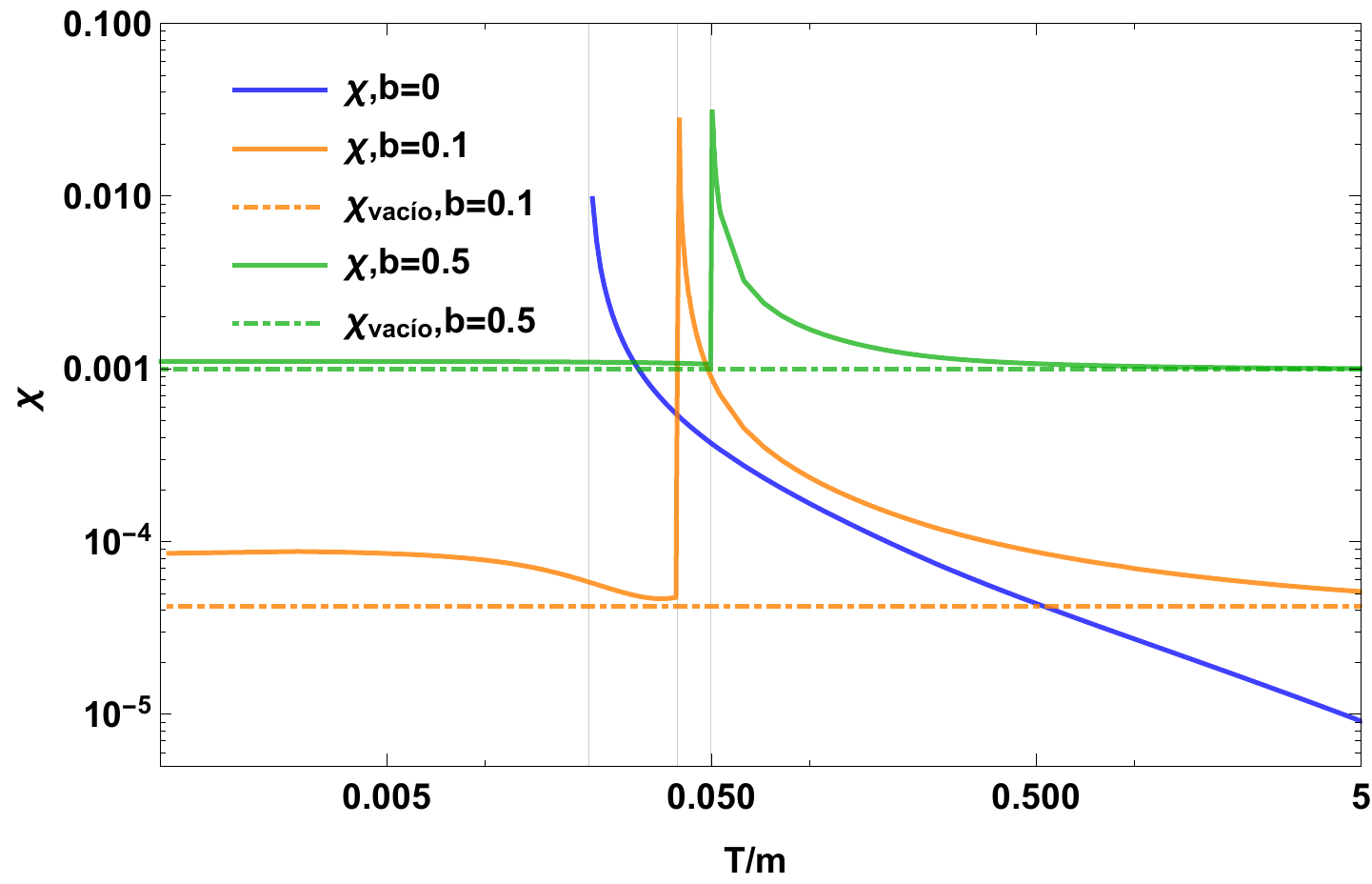}
	\caption{\label{f3.31} The specific heat (left) and the magnetic susceptibility (right) as a function of $b$ and $T$ for $ \rho=1.30 \times 10^{39}$cm$^{-3} $. The vertical lines signal the temperature of condensation.}
\end{figure}

Fig.~\ref{f3.31} shows the specific heat and the magnetic susceptibility as a function of temperature for $ \rho= 1.30 \times 10^{39}$ cm$^{-3} $ and several values of the magnetic field. As in the non-relativistic case \cite{Lismary2018}, the peaks of both magnitudes occur at the condensation temperature (the solid vertical lines). This reinforces our conclusion that the magnetic behavior of the gas below $T_c$ is a consequence of condensation. From Eqs.~(\ref{chi}-\ref{chiTc}) follows that at $b=0$, $\chi$ diverges for all $ T <T_c $. The latter was also obtained in \cite{yamada1982thermal,Lismary2018} for the non-relativistic susceptibility and constitutes another evidence of the relation between BEC and the magnetic properties of the gas.

The appearance of a spontaneous magnetization and a zero-field divergent magnetic susceptibility below the critical BEC temperature have been found in \cite{PhysRevD.52.7174,ROJAS1996148,Khalilov1997,PEREZROJAS2000,Khalilov:1999,Jian,yamada1982thermal,Simkin_1999,Lismary2018} for charged and non-relativistic neutral vector bosons under the action of an external magnetic field. Besides the similarities, we would like to remark that Bose-Einstein ferromagnetism is not true ferromagnetism since its cause is the combination of Bose-Einstein statistics with a ground state that only contains particles with $s=1$, and not spin-spin interactions like in conventional ferromagnets. Nevertheless, in experimental situations with real gases, spin-spin interactions, although weak, could also contribute to the magnetic properties of vector bosons \cite{Simkin_1999}.

Another remarkable feature of Fig.~\ref{f3.31} is the high temperature magnetic susceptibility. When $ b = 0 $, $\chi$ decreases with  $T$. But at finite magnetic field, $\chi \rightarrow \chi_{vac}$ when $ T $ increases. This is consistent with the fact that the magnetization augments with the temperature rather than canceling out. As we have already seen, this is a consequence of the presence of a finite fraction of antiparticles in the system and a main difference with respect to the NR limit. The effect of antiparticles is also present in the specific heat, that for high temperature increases instead of tending to the classical value $3/2$.

\section{Anisotropic pressures}
\label{sec6}

Now we analyze how antiparticles and magnetic field affects the parallel $P_{\parallel}=-\Omega$ and perpendicular $P_{\perp}=-\Omega-\mathcal M B$ pressures of the gas. According to their definition, $P_{\parallel}$ and $P_{\perp}$ are the spatial components of the statistical average of the energy momentum tensor of the system of bosons under the action of an uniform and constant external magnetic field \cite{Quintero2017AN,Ferrer:2010wz,Bali:2014kia}. In this context ``parallel'' and ``perpendicular'' is said with respect to the magnetic field direction. This anisotropy in the pressure is important for the the gravitational stability of astronomical objects \cite{ManrezaParet:2020poe}, and is also connected with an interesting phenomenon known as quantum magnetic collapse \cite{Chaichian:1999gd}.
\begin{figure}[ht!]
	\centering
	\includegraphics[width=0.49\linewidth]{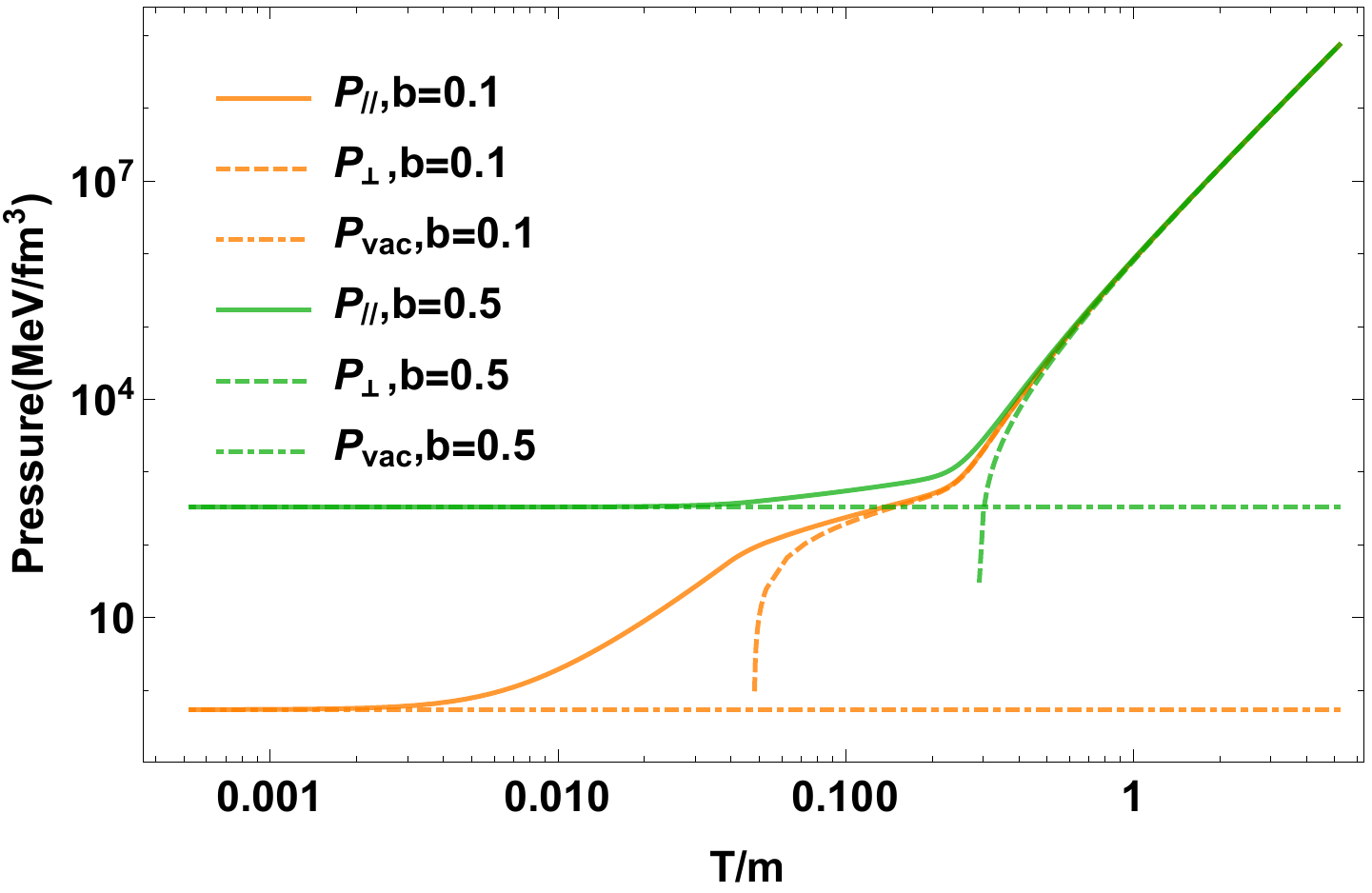}
	\caption{\label{f3.3} The pressures as functions of the temperature and the magnetic field for $ \rho =1.30 \times 10^{39}$ cm$^{-3}$.}
\end{figure}

Fig.~\ref{f3.3} shows the pressures vs the temperature for $ \rho = 1.30 \times 10^{39} $cm$^{-3} $ and various values of the magnetic field. The vacuum pressure $P_{vac}=-\Omega_{vac}$ is also drawn for comparison. In this plot two regions can be clearly identified: in the first one, $ T> m $ and temperature dominates; therefore, the difference between the pressures is negligible. On the contrary, in the second one $T<m$, the magnetic field dominates, and the presence of the magnetized vacuum in $P_{\parallel}$ and the term $-\mathcal M B$ in $P_{\perp}$ are apparent. As $T \rightarrow 0$ the difference between the pressures increases; $P_{\parallel}$ tends to the constant value $-\Omega_{vac}(b)$, while $P_{\perp}$ becomes negative at the point at which $-\Omega = \mathcal M B$. (Let us recall that for a Bose gas the statistical part of the pressure $-\Omega_{st}$ goes to zero with temperature.)

\begin{figure}[ht!]
	\centering
	\includegraphics[width=0.49\linewidth]{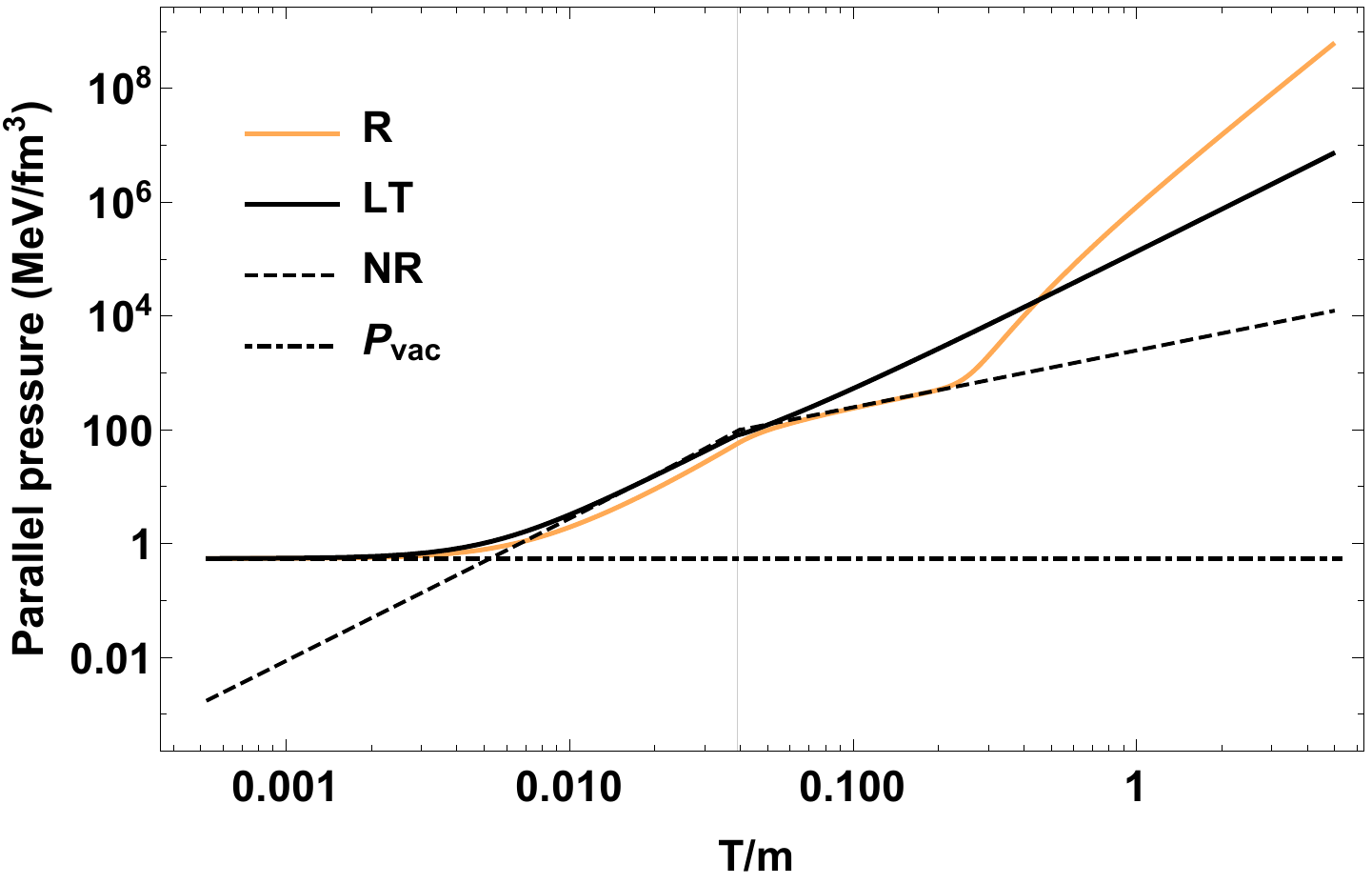}
	\includegraphics[width=0.49\linewidth]{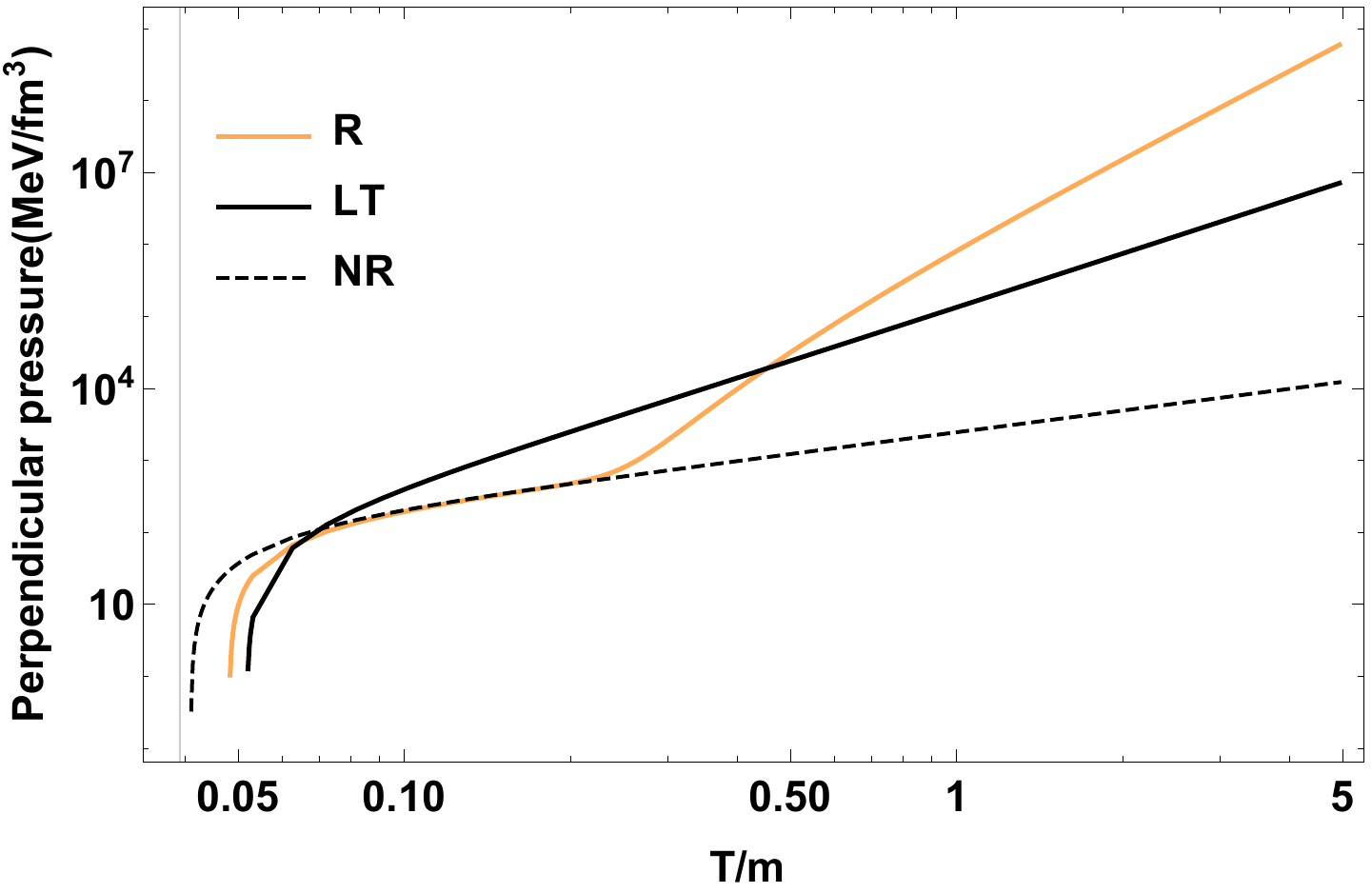}	
	\caption{\label{f3.4} The R, NR and LT parallel and perpendicular pressures as functions of temperature for $ \rho =1.30 \times 10^{39}$ cm$^{-3} $ and $ b=0.1$.}
\end{figure}

The differences between the pressures resulting from the relativistic calculation at all temperature and their non-relativistic and low temperature counterparts are shown in Fig.~\ref{f3.4}. They are three: first, the presence of antiparticles in the region of high temperatures cause a difference of several orders between the R and the NR pressures; second, in the relativistic cases the parallel pressure at low temperatures is dominated by $-\Omega_{vac}(b)$, while in the NR limit $ P_{\parallel} $ tends to zero with $ T $; and third, the value of temperature where $ P_{\perp} = 0 $ is underestimated in the NR limit and overestimated in the LT approximation.
Note that below the temperature at which $ P_{\perp} = 0 $, the perpendicular pressure becomes negative and the gas is unstable. Such instability is known as quantum magnetic collapse \cite{Chaichian:1999gd}. It seems to suggest that, for a fixed temperature, the magnetic field presence imposes an upper bound on the boson's density needed to sustain it, but this point worth further comment. To model astrophysical environments, in particular, when considering the gravitational equilibrium of astrophysical bodies, Maxwell´s contribution to the pressures $B^2/4 \pi$ should be taken into account \cite{Broderick_2000}. It modifies the pressures in the following form: $P_{\parallel}\rightarrow P_{\parallel} +B^2/4 \pi $ and $P_{\perp}\rightarrow P_{\perp} - B^2/4 \pi $. Thus, their relation flips from $P_{\parallel}>P_{\perp}$ with no Maxwell contribution to $P_{\parallel}<P_{\perp}$ with Maxwell contribution \cite{Quintero2018BECS}. In this situation the quantum magnetic collapse is still possible, but now the instability occurs for $P_{\parallel}$ instead of $P_{\perp}$ \cite{Quintero2018BECS}. Nevertheless, in systems composed of many kinds of particles, the effect of the Maxwell term might be balanced by the pressures of some of the gases, being the relation $P_{\parallel}>P_{\perp}$ and the possibility of the $P_{\perp}<0$ instability recovered \cite{2021IJMPD..3050007F}.

\section{Concluding remarks}
\label{sec}

We computed the exact analytic expressions for the thermodynamic quantities of a relativistic magnetized neutral vector boson gas at any temperature, including magnetization and the second derivatives of the thermodynamical potential ($C_v$ and $\chi$). Our calculations were inspired by astrophysics, however, they have a \emph{per se} interest and can be useful in other scenarios like particle physics and condensed matter physics.

The numerical study of the magnetized NVBG in astrophysical conditions allowed us to evaluate the relative influence of the particle density, the magnetic field and temperature in the system. Depending on $T$ there are two distinct regimes in the behavior of the gas. For $ T << m $ the effects of the magnetic field dominate the system and, in particular, we checked its relevant role in Bose--Einstein condensation, pressure anisotropy and quantum magnetic collapse.

When $ T >> m $ the temperature effects dominate, being the most important, the existence of a non-negligible fraction of antiparticles. In general, the density of antiparticles is no longer negligible around $ T \sim 0.25 \; m $, although their effects are most strongly manifested for $ T \gtrsim m $, being both values of temperature quite independent of the magnetic field. The relevance of antiparticles is specially evident for the magnetization and the pressure of the gas, since they cause an increase in various orders in both magnitudes. 

When $ B \rightarrow 0 $ the NVBG remains magnetized, as in the NR and the LT limits. This spontaneous magnetization is not associated with an interaction between the spins of the bosons, but it is a consequence of condensation. The ability of vector gases to spontaneously magnetize may be connected with the origins of magnetic fields in astrophysical environments.

The comparison of the all temperature relativistic calculations with its NR and LT limits allowed to established the validity ranges of these approximations and the physics they ignore. In the NR limit, increasing the temperature means going to the classical case, i.e. it is equivalent to having the Boltzmann distribution function for the particles. However, when we increase $T$ in the relativistic case, the antiparticles plays the main role, contributing to all the magnitudes and introducing non trivial differences between both situations. Something similar happens with the vacuum pressure and magnetization: they are usually neglected, however, they effects are important for high magnetic field. Therefore, the NR limit is valid only for low temperatures and weak magnetic fields.

The LT limit, on the contrary, works well for relatively strong magnetic fields $ B \geq 0.2 B_c $. Looking at some fixed values we find that, for example, for $ B =10^{16} $ G, this approximation is valid for temperatures such that $ T << 10^{- 3} m $. This implies that for paired neutrons this limit cannot be used for $ T \sim10^{10}$ K, which is a possible temperature in early stages of NS.  In addition, we found that increasing $T$ in the LT limit leads to a negative magnetization, while our all temperature study shows that the magnetization of the gas is always positive. Hence the importance of using the exact expressions even for low temperatures. 

\section{Acknowledgments}
We thank E. Mart\'inez Rom\'an and C. Reigosa Soler for their review and useful comments on the manuscript. The authors have been supported by the grant No.~500.03401 of the PNCB-MES, Cuba, and the grant of the Office of External Activities of the Abdus Salam International Centre for Theoretical Physics (ICTP) through NT-09. 

\appendix

\section{Vacuum thermodynamic potential}\label{appA}

Here we compute the renormalized vacuum contribution to the thermodynamic potential Eq.(\ref{Grand-Potential-vac}) following \cite{Quintero2017PRC}. We start from $\Omega_{vac}$ definition
\begin{equation}\label{Grand-potential-vac-1}
\Omega_{vac}=\sum_{s=-1,0,1}\int\limits_{0}^{\infty}\frac{p_{\perp}dp_{\perp}dp_3}{(2\pi)^2}\varepsilon (p_{\perp},p_3, B,s) \nonumber
\end{equation}
with
\begin{equation}\nonumber
\varepsilon(p_{\perp},p_3, B,s)=\sqrt{p_3^2+p_{\perp}^2+m^2-2\kappa s B\sqrt{p_{\perp}^2+m^2}}.
\end{equation}
We integrate over $p_3$ and $p_{\perp}$ with the help of the equivalence
\begin{equation}
\sqrt{a}= -\frac{1}{2 \sqrt{\pi}} \int\limits_{0}^{\infty} dy y^{-3/2} (e^{- y a}-1)
\end{equation}
\noindent and the introduction of the small quantity $\delta$ as lower limit of the integral
\begin{equation}
\sqrt{a}= -\frac{1}{2 \sqrt{\pi}} \int\limits_{\delta}^{\infty} dy y^{-3/2} e^{-y a}.
\end{equation}
The latter is done to regularize the divergence of the $a$ dependent term and to eliminate the term that does not depends on $a$.

Now, let's make $a = \varepsilon^2 = p_3^2+p_{\perp}^2+m^2-2\kappa s B\sqrt{p_{\perp}^2+m^2}$. As a consequence
\begin{equation}\label{energyintegral}
\varepsilon = -\frac{1}{2 \sqrt{\pi}} \int\limits_{\delta}^{\infty} dy y^{-3/2} e^{- y(p_3^2+p_{\perp}^2+m^2-2\kappa s B\sqrt{p_{\perp}^2+m^2})},
\end{equation}
where we dropped out the explicit dependence of the spectrum on $p_{\perp},p_3, B$ and $s$ to simplify the writing.
Inserting Eq.(\ref{energyintegral}) in Eq.(\ref{Grand-potential-vac-1}), the vacuum thermodynamic potential reads as follows
\begin{eqnarray}\label{Grand-potential-vac-2}
\Omega_{vac}=-\frac{1}{8 \pi^{5/2}}\sum_{s=-1,0,1}\int\limits_{\delta}^{\infty} dy y^{-3/2} \int\limits_{0}^{\infty}dp_{\perp} p_{\perp}
 \int\limits_{-\infty}^{\infty}dp_3 e^{- y(p_3^2+p_{\perp}^2+m^2-2\kappa s B\sqrt{p_{\perp}^2+m^2})}.
\end{eqnarray}
After integration over $p_3$ we obtain
\begin{eqnarray}\label{Grand-potential-vac-3}
\Omega_{vac}=-\frac{1}{8 \pi^{2}}\sum_{s=-1,0,1}\int\limits_{\delta}^{\infty} dy y^{-2} \int\limits_{0}^{\infty}dp_{\perp} p_{\perp} e^{- y(p_{\perp}^2+m^2-2\kappa s B\sqrt{p_{\perp}^2+m^2})}.
\end{eqnarray}
Eq.(\ref{Grand-potential-vac-3}) may be simplified by two successive changes of variables. The first one is $z = \sqrt{m^2+p_{\perp}^2} - s \kappa B$, and Eq.(\ref{Grand-potential-vac-3}) becomes
\begin{eqnarray}\label{Grand-potential-vac-4}
\Omega_{vac}=-\frac{1}{8 \pi^{2}}\sum_{s=-1,0,1} \left \{\int\limits_{\delta}^{\infty} dy y^{-3} e^{- y(m^2-2 m s \kappa B)}
+ s \kappa B \int\limits_{\delta}^{\infty} dy y^{-2} \int\limits_{z_1}^{\infty}dz e^{- y(z^2 - s^2 \kappa^2 B^2)}
\right \},
\end{eqnarray}
\noindent where $z_1 = m - s \kappa B $.
The second change of variables is $w=z-z_1$ in the last term of Eq.(\ref{Grand-potential-vac-4}). If, in addition, we sum over the spin and recall that $b = B/B_c$ with $B_c = m / 2 \kappa$, $\Omega_{vac}$ can be written as
\begin{eqnarray}\label{Grand-potential-vac-5}
\Omega_{vac} &=&-\frac{1}{8 \pi^{2}}\left \{ \int\limits_{\delta}^{\infty} dy y^{-3} e^{- ym^2} (1+2 \cosh{[m^2 b y]}) \right. \\\nonumber 
 &+&  m b \left. \int\limits_{\delta}^{\infty} dy y^{-2} \int\limits_{0}^{\infty} dw e^{- y(m - w)^2} \sinh[m b (m - w) y] \right\}. 
\end{eqnarray}
We can identify the ultraviolet divergencies related to the terms of Eq.~(\ref{Grand-potential-vac-5}) proportional to ($1+2 \cosh{[m^2 b y]}$) and $ \sinh[m b (m - w) y]$ by expanding them in powers of $ y b$ up to the higher divergent term
\begin{eqnarray}\label{Grand-potential-R-1}
	\Omega^{div}_{vac} &=& -\frac{1}{8 \pi^{2}}\int\limits_{0}^{\infty} dy y^{-3} e^{- ym^2} (3 + m^4 b^2 y^2)\\
	&-&\frac{m b}{8 \pi^{2}} \int\limits_{0}^{\infty} dy y^{-2} \int\limits_{0}^{\infty}dw e^{- y(m - w)^2}
	\left \{ m b (m - w) y + [m b (m - w) y]^3/6 \right \}. \nonumber
\end{eqnarray}
Now, to take the limit $\delta \rightarrow 0$ and eliminate the divergent part of $\Omega_{vac}$, we add and subtract Eq.~(\ref{Grand-potential-R-1}) to Eq.~(\ref{Grand-potential-vac-5}). 

To interpret in an easy way the divergent terms added to Eq.~(\ref{Grand-potential-vac-5}), we simplify Eq.~(\ref{Grand-potential-R-1}) through the following opperations: integration over $w$, the change of variables $x=y m^2$ and the substitution $b=B/B_c=2 \kappa B/m$. After those changes Eq.~(\ref{Grand-potential-R-1}) reads 
\begin{eqnarray}\label{Grand-potential-R-3}
	\Omega_{vac}^{div} &=& -\frac{m^4}{8 \pi^{2}}\int\limits_{0}^{\infty} dx \frac{e^{- x}}{x^3}
	-\frac{ (m \kappa B)^2}{2 \pi^{2}} \int\limits_{0}^{\infty} dx e^{-x}\frac{2x-1}{2 x^2}  + \frac{ (\kappa B)^4}{6 \pi^{2}} \int\limits_{0}^{\infty} dx e^{-x} \frac{1+x}{x}.
\end{eqnarray}
The first divergent term of Eq.~(\ref{Grand-potential-R-3}) is the value of $\Omega_{vac} (B=0)$. Thus, it incorporates into the zero-point energy of the magnetized system. The second divergent term is absorbed by the bare magnetic field energy. It redefines -renormalizes- the magnetic field and the coupling constant of the bosons. Last term also incorporates to the zero-point energy that, for this magnetized gas, depends on $\kappa$ and $B$ (a similar term appears for magnetized fermions with anomalous magnetic moment \cite{Ferrer_2015}). 

Finally, the renormalized vacuum potential  is given by
\begin{eqnarray}\label{Grand-potential-vac-6}
\Omega_{vac} &=& -\frac{1}{8 \pi^{2}}\int\limits_{0}^{\infty} dy y^{-3} e^{- ym^2} \{2 \cosh{[m^2 b y]} - 2 - m^4 b^2 y^2 \}\\
&-&\frac{m b}{8 \pi^{2}} \int\limits_{0}^{\infty} dy y^{-2} \int\limits_{0}^{\infty}dw e^{- y(m - w)^2}
\left \{ \sinh[m b (m - w) y] \right.\nonumber\\ 
&-& \left. m b (m - w) y - [m b (m - w) y]^3/6 \right \}. \nonumber
\end{eqnarray}
After integration Eq.(\ref{Grand-potential-vac-6}) leads to
\begin{align}\nonumber
	\Omega_{vac}(b) = -\frac{m^4}{288 \pi}\left( b^2(66-5 b^2)
	-3(6-2b-b^2)(1-b)^2 \log(1-b)\right.
	\\\nonumber
	\left. -3(6+2b-b^2)(1+b)^2\log(1+b) \right)
\end{align}
which is Eq.~(\ref{Grand-Potential-vac}) of the main text.

\section{Thermodynamic potential of the NVBG in the low temperature limit}\label{appB}

To obtain the low temperature limit of the statistical thermodynamic potential Eq.~(\ref{Grand-Potential-sst21}), we transform Eq.~(\ref{Grand-Potential-sst2}) by computing the integral on its second term \cite{Quintero2017AN}
\begin{eqnarray}\label{I3}
I = \int\limits_{y_0}^{\infty} dz
\frac{x^2}{\sqrt{x^2+\alpha^2}} K_1 (n \beta x).
\end{eqnarray}

Let's introduce the following form for $K_1 (n \beta x)$
\begin{equation}\label{k1int}
K_1 (n \beta x) = \frac{1}{n \beta x}\int\limits_{0}^{\infty} dt e^{-t-\frac{n^2 \beta^2 x^2}{4 t}}.
\end{equation}

If we substitute (\ref{k1int}) in (\ref{I3}), the integration over $x$ can be carried out
\begin{equation}\label{I31}
I= \frac{\sqrt{\pi}}{n^2 \beta^2} \int\limits_{0}^{\infty} dt \sqrt{t} e^{-t+\frac{n^2 \beta^2 \alpha^2}{4 t}} erfc \left(\frac{n \beta \sqrt{y_0^2+\alpha^2}}{2 \sqrt{t}}\right).
\end{equation}

To integrate over $t$ in (\ref{I31}) we replace the complementary error function $erfc(x)$ by its series expansion
\begin{equation}\label{erfc}
erfc(x) \backsimeq \frac{e^{-x^2}}{\sqrt{\pi} x} \left(1 - \sum_{w=1}^{\infty} \frac{(-1)^w(2 w -1)!!}{(2 x^2)^w}\right).
\end{equation}
After integration Eq.~(\ref{I31}) becomes
\begin{eqnarray}\label{I32}
\hspace{-.5cm} I = \frac{z_0^2}{n \beta \sqrt{y_{0}^2 + \alpha^2}} K_2 (n \beta y_0) - \hspace{-.5cm}
\frac{y_0^2}{n \beta \sqrt{y_{0}^2 + \alpha^2}} \sum_{w=1}^{\infty} \frac{(-1)^w(2 w -1)!!}{(y_0^2+\alpha^2)^w} \left(\frac{y_0}{n \beta}\right)^w \hspace{-.2cm} K_{-(w+2)} (n \beta y_0).
\end{eqnarray}

By sustituting Eq.~(\ref{I32}) in Eq.~(\ref{Grand-Potential-sst2}), $\Omega_{st}(s)$ reads
\begin{eqnarray}\label{Grand-Potential-sst3}
\Omega_{st}(s)&=& - \frac{y_0^2 }{2 \pi^2 \beta^2} \left(1+\frac{\alpha}{\sqrt{z_0^2 + \alpha^2}}\right)  \sum_{n=1}^{\infty} \frac{e^{n \mu \beta}+e^{- n \mu \beta }}{n^2}  K_2 (n \beta y_0)\\\nonumber
&-& \frac{\alpha y_0^2}{ \pi^2 \beta^2 \sqrt{y_{0}^2 + \alpha^2}} \sum_{n=1}^{\infty} \frac{e^{n \mu \beta}+e^{- n \mu \beta }}{n^2}
 \sum_{w=1}^{\infty} \frac{(-1)^w(2 w -1)!!}{(y_{0}^2 + \alpha^2)^w} \left(\frac{y_0}{n \beta}\right)^w K_{-(w+2)}(n \beta y_0).
\end{eqnarray}

Taking the low temperature limit $T \ll m$ in Eq.~(\ref{Grand-Potential-sst3}) is equivalent to make $\beta \rightarrow \infty$. In this limit all the terms in $\Omega_{st}(s)$ goes to zero except for the first one, therefore
\begin{eqnarray}
\Omega_{st}(s) \cong - \frac{y_0^2}{2 \pi^2 \beta^2} \left(1+\frac{\alpha}{\sqrt{y_0^2 + \alpha^2}}\right) \sum_{n=1}^{\infty} \frac{e^{n \mu \beta}+e^{- n \mu \beta }}{n^2} K_2 (n \beta y_0).
\end{eqnarray}

In addition
\begin{equation}
K_2 (n \beta y_0) \cong \frac{\sqrt{\pi} e^{-n \beta y_0}}{\sqrt{2 n \beta y_0}} = \frac{\sqrt{\pi} e^{-n \beta y_0}}{\sqrt{2 n \beta y_0}},\nonumber
\end{equation}
\noindent and $\Omega_{st}(s)$ can be written as
\begin{eqnarray}\nonumber
\Omega_{st}(s) \cong - \frac{y_0^{3/2}}{2^{3/2} \pi^{3/2} \beta^{5/2}} \left(1+\frac{\alpha}{\sqrt{y_0^2 + \alpha^2}}\right) \sum_{n=1}^{\infty} \frac{e^{n \beta (\mu - y_0)}+e^{- n \beta (\mu + y_0) }}{n^{5/2}}.
\end{eqnarray}

For $\beta \gg 1$, the antiparticles term $e^{- n \beta (\mu + z_0)}$ goes to zero for all spin eigenvalues $s = 0, \pm 1$, because $\mu + y_0$ ($y_0= m \sqrt{1-s b}$) is always a positive quantity. So, in the low temperature limit the antiparticles contribution can be neglected.

Similarly, the particle term $e^{ n \beta (\mu - z_0)}$ goes to zero for $s = 0, -1$, since in these cases $\mu - y_0$ is negative. However, when $s=1$, $e^{ n \beta (\mu - z_0)}$ goes to $1$, because $\mu \rightarrow m\sqrt{1-b}=y_0(s=1)$. As a consequence, for low enough temperatures, the contribution of the particles with spin states  $s=0,-1$ are also negligible and $\Omega_{st} \cong \Omega_{st}(1)$. Finally, since $\Omega_{st}(1)$ admits further simplifications, the statistical part of the thermodynamical potential in the low temperature limit is equal to
\begin{eqnarray}\label{Grand-Potential-stfinal}
\Omega^{LT}_{st}(b,\mu,T) =  -\frac{\left(m \sqrt{1-b}\right)^{3/2}}{2^{1/2} \pi^{3/2} \beta^{5/2} (2-b)} Li_{5/2}(e^{\beta \mu^{\prime}}),
\end{eqnarray}

\noindent where  $Li_{n}(x)=\sum_{l=1}^\infty x^l/l^{n}$ is the polylogarithmic function of order $n$ and $ \mu^{\prime} = \mu - m \sqrt{1-b}$.

Using Eq.~(\ref{Grand-Potential-stfinal}) instead of Eq.~(\ref{Grand-Potential-sst21}) in Eq.~(\ref{omega}) the thermodynamic magnitudes are computed in the relativistic low temperature limit \cite{Quintero2017PRC}. They read
\begin{subequations}
	\begin{align}\label{EoSNB1}
	\rho^{LT}&= \rho^{LT}_{gs} +\frac{\varepsilon^{3/2} Li_{3/2}(e^{\mu^\prime \beta})}{\sqrt{2\pi}\, \pi \beta^{3/2} (2-b)}\\[2mm]
	\mathcal M^{LT}_{st}&= \frac{\kappa}{\sqrt{1-b}} \rho^{LT},\\[2mm]
	P^{LT}_{\parallel} &= -\Omega^{LT}_{st} -\Omega_{vac},\\[2mm]
	P^{LT}_{\perp} &= -\Omega^{LT}_{st} -\Omega_{vac}-\mathcal M^{LT} B,\\[2mm]
		E^{LT} &= m\sqrt{1-b} \rho^{LT} + \Omega_{vac}- \frac{3}{2} \Omega^{LT}_{st},
	\end{align}
\end{subequations}
with $\rho^{LT}_{gs}=\rho\left[1-(T/T^{LT}_c)^{3/2}\right]$ the density of condensed particles and
\begin{eqnarray}\label{Tclt}
\mu^{\prime} &=& -\frac{\zeta(3/2)T}{4 \pi} \left [ 1- \left(\frac{T^{LT}_{c}}{T} \right)^{3/2} \right ]\Theta(T-T^{LT}_{c}),\\[2mm]
T^{LT}_{c} &=& \frac{1}{m\sqrt{1-b}} \left [ \frac{\sqrt{2\pi}\, \pi (2-b) \rho^{LT}}{\zeta(3/2)}\right]^{2/3},
\end{eqnarray}
where $T^{LT}_c$ is the LT critical temperature of condensation and $\zeta(x)$ is the Riemann zeta function.

\section{Thermodynamic potential in the non relativistic limit}\label{appC}

In this appendix we compute the thermodynamic potential of the magnetized vector boson gas in the non--relativistic limit as done in \cite{Lismary2018}. We start from the non-relativistic spectrum $\varepsilon(p,s)=\vec{p}^{\:2}/2m -s\kappa B$ and consider the density of states of the gas, that is
\begin{eqnarray}\nonumber
g(\epsilon)&=&\frac{4\pi V}{(2\pi \hbar)^3}
\sum_{s=-1,0,1}  \int_0^\infty dp\;p^2 \delta\left(\epsilon-\frac{\vec{p}^{\:2}}{2m}+ s\kappa B\right),
\end{eqnarray}
\noindent where $\epsilon$ is the boson energy. Let us note that the rigorous no relativistic limit obtained from the spectrum Eq.~(\ref{spectrum}) in the NR limit $p_3, p_{\perp}, \kappa B << m$ is
\begin{equation}
\varepsilon(p,s)=m+\vec{p}^{\:2}/2m -s\kappa B.
\end{equation}

\noindent However, to simplify the calculations, we have done the rescaling $\varepsilon \rightarrow \varepsilon-m$. This is equivalent to do the substitution $\mu \rightarrow \mu-m$ in the thermodynamic potential, and the only magnitude affected is the energy density, but it can be easily corrected by the addition of $m \rho$.

After doing the integration over $p$ and the sum over the spin states $s$, $g(\epsilon)$ becomes
\begin{eqnarray}\label{eq3}
g(\epsilon)=  \frac{4\pi mV}{(2\pi \hbar)^3} \left[ \sqrt{2m(\epsilon-\kappa B)}+ \sqrt{2m \epsilon} + \sqrt{2m(\epsilon +\kappa B)} \right].
\end{eqnarray}

Note that Eq.~(\ref{eq3}) can be separated in three terms, each one corresponding to a specific spin state. Since the thermodynamical potential $\Omega^{NR}(\mu,T,B)$ is
\begin{equation}\label{eq4}
\Omega^{NR}(\mu,T,B)=\frac{T}{V}\int_0^\infty d\epsilon g(\epsilon)\;ln (\emph{f}_{BE} (\epsilon,\mu)), \;\;\forall\;\mu<\epsilon,
\end{equation}
\noindent with $\emph{f}_{BE}(\epsilon,\mu)=\left[1-e^{\beta(\mu-\epsilon)}\right]^{-1}$, it can also be separated in three terms $\Omega^{NR}(\mu,T,B)=\Omega_{-}(\mu,T,B)+\Omega_{0}(\mu,T,B)+\Omega_{+}(\mu,T,B)$, corresponding to the states with $s=-1$, $s=0$ and $s=1$ respectively. Integrating over the energy in Eq.~(\ref{eq4}) one gets

\begin{eqnarray}\label{eq12}
\Omega_{-}(\mu,T,B)&=&- \frac{T}{\lambda^3} Li_{5/2}(z_{-}),\\
\Omega_{0}(\mu,T,B)&=&- \frac{T}{\lambda^3} Li_{5/2}(z),\label{eq1a2} \\
\Omega_{+}(\mu,T,B)&=& - \frac{T}{\lambda^3} Li_{5/2}(z_{+}),\label{eq12b}
\end{eqnarray}

\noindent where $\lambda=\sqrt{2\pi/mT}$ is the thermal wavelength, $z=e^{\mu/T}$ is the fugacity and $z_{\sigma}=z e^{\sigma \frac{\kappa B}{T}}$ where $\sigma=-,+$.

Eqs.~(\ref{eq12}) and (\ref{eq12b}) allow to compute all the thermodynamic magnitudes of the non--relativistic neutral vector boson gas. In particular, one has the following expressions for the particle density

\begin{eqnarray}\label{eq18}
\rho^{NR} &=&\rho^{NR}_{gs}(T,B) + \rho_{-}(\mu,T,B)+\rho_{0}(\mu,T,B)+\rho_{+}(\mu,T,B),\nonumber\\
\end{eqnarray}
\noindent where $\rho_{gs}$ stands for the particles in the condensate and
\begin{eqnarray}
\rho_{-}(\mu,T,B)&=&\frac{Li_{3/2}(z_{-}) }{\lambda^3},\nonumber\\
\rho_{0}(\mu,T,B)&=&\frac{Li_{3/2}(z)}{\lambda^3},\nonumber\\
\rho_{+}(\mu,T,B)&=&\frac{Li_{3/2}(z_{+})}{\lambda^3},\nonumber
\end{eqnarray}
\noindent correspond to the density of particles with spin state $-1$, $0$ y $1$ respectively; the magnetization
\begin{equation}\label{eq28}
\mathcal M^{NR}= \kappa\rho^{NR}_{gs} -\left(\frac{\partial \Omega^{NR}}{\partial B}\right) =  \kappa(\rho^{NR}_{gs} + \rho_{+})-\kappa \rho_{-};
\end{equation}
\noindent and the energy and pressures
\begin{eqnarray}\label{EoS}
E^{NR} &=&\frac{3}{2} P^{NR}_{\parallel} - \kappa B (\rho^{NR}_{gs} + \rho_{+} - \rho_{-}), \label{energia}\\
P^{NR}_{\parallel} &=&  -\Omega^{NR}, \label{presionpara} \\
P^{NR}_{\perp}&=& -\Omega^{NR} -{\mathcal M^{NR}} B. \label{presionper}
\end{eqnarray}


%

\end{document}